\newcommand{\pr}{\mathbb{P}}	
\newcommand{\E}{\mathbb{E}}	    

\documentclass[final,3p,times,authoryear]{elsarticle}

\usepackage{amssymb}
\usepackage{amsmath}

\journal{Infectious Disease Modelling}

\usepackage{lscape}
\usepackage{booktabs} 
\usepackage{xcolor}
\usepackage{url}

\newif\ifclean
\cleantrue 

\ifclean
  \renewcommand{\textcolor}[2]{#2}
\fi

\begin{document}
\let\d\displaystyle 

\begin{frontmatter}


\author{Jordi Ripoll\corref{cor1}} 
\ead{jordi.ripoll@udg.edu}
\author{Joan Saldaña}
\ead{joan.saldana@udg.edu}
\cortext[cor1]{corresponding author}
\affiliation{organization={Department of Computer Science, Applied Mathematics and Statistics}, 
            addressline={Campus Montilivi, 17003}, 
            city={Girona},
            country={Spain}}

\title{Generation time in a discrete epidemic model with asymptomatic carriers: beyond geometric waiting times} 




\begin{abstract}
We study the random times between successive cases in a transmission chain of infectious diseases with asymptomatic carriers. We derive the probability distribution of this generation time (in days) from a discrete-time epidemic model with variable infectiousness both along elapsed times and across phases. The introduced non-Markovian model is a compact recursive system featuring random waiting times at each of the three infected stages: latent, asymptomatic, and symptomatic. By rearranging the terms of the basic reproduction number, which represents the expected number of secondary cases produced by an asymptomatic primary case who may eventually develop symptoms, we get to the generation-time probabilities. The expected generation time is a convex combination of the expected generation times before and after the onset of symptoms. Additionally, our analysis reveals that the $n$-th moment of the generation time is related to the moments up to $n$-th order of the weighted forward recurrence time at each phase and the moments up to $n$-th order of the latent period and the incubation period. These weights are the infectiousness along the elapsed times for each transmission phase. Finally, we illustrate several data-driven epidemic scenarios, assuming that infectiousness varies only across phases and discrete Weibull distributions for the waiting times. \textcolor{blue}{Each disease analyzed shows a range of variability in its generation time distribution, from low to moderate.} 
\end{abstract}




\begin{keyword}
Discrete-time epidemic model\sep asymptomatic transmission\sep variable infectiousness \sep renewal equation\sep basic reproduction number \sep discrete Weibull distribution \sep residual time


\end{keyword}

\end{frontmatter}



\section{Introduction}

Most traditional compartmental epidemic models assume that the rates of progression, transmission, and recovery are constant. In continuous time, this results in an exponential distribution for the latent and infectious periods; in discrete time, it amounts to a geometric distribution of these periods. While this is a mathematically convenient assumption, it is known that it does not reflect real-world observations. For instance, studies of the transmission dynamics of measles in small communities have shown that the empirical distribution of infectious periods is less dispersed than that corresponding to an exponential distribution \citep{Lloyd(2001)}. For other infectious diseases like Ebola or SARS-CoV-2, the empirical distributions have been approximated by gamma and Weibull distributions \citep{Chen(2022),Chowell(2014),Ferretti(2020)}.   

The distributions of these sojourn times have an impact on the generation time distribution (the time from the infection of a primary case to infection of the cases it generates), which, in turn, is used to estimate the basic reproduction number from the exponential growth of the number of cases during the initial phase of an epidemic \citep{Wallinga(2007)}. Moreover, when infectious period distributions are more homogeneous than the exponential/geometric one, stochastic models of childhood diseases such as measles predict smaller critical community sizes and generate low-amplitude outbreaks (``generation pulses'') that are consistent with observations from small communities \citep{Keeling(1997)}. Therefore, it is important to incorporate more realistic aspects of the disease transmission into models, such as the fact that an individual's infectiousness varies throughout his/her infectious period depending on the time since infection, the so-called age of infection. This dependence is caused by factors such as a varying viral load or changes in the contact patterns of infected individuals. 

Most age-of-infection models have been formulated in continuous time and using partial differential equations to describe how the density of infected individuals changes over time \citep{Bai(2023)}.  However, discrete-time epidemic models are also a natural framework for incorporating dependence on the age of infection or other elapsed times \citep{Hernandez(2013)}. This flexibility, their straightforward numerical implementation, alongside the fact that the surveillance data are often reported as daily or weekly case counts (i.e., discrete counts), explain the renewed interest in discrete epidemic models \citep{Elaydi(2025),Diekmann(2021),Sanz(2024),Thieme(2024)}.  

One of the features highlighted during the COVID-19 pandemic was the important role played by asymptomatic individuals in transmitting the disease. Asymptomatic infection had already been observed in other infectious diseases, such as Ebola \citep{Chowell(2014)}. To deal with this fact, we will consider a discrete-time model with two possible outcomes for asymptomatic cases: they either progress to showing symptoms (pre-symptomatic cases) or recover asymptomatically (persistent asymptomatic cases) \citep{Fraser(2004)}. Paraphrasing the book \citep{Weitz(2024)}, asymptomatic spread is a silent or mild outcome for some that can lead, inadvertently, to far greater numbers of severe cases for the population as a whole.

In this paper, we formulate a Susceptible-Exposed-Asymptomatic-Recovered/Infected-Deceased (SEA-RID) epidemic model. To avoid complex mathematical formulations, the non-Markovian structured model builds upon the Markovian recursive system by embedding elapsed times into both the state variables and parameters. This enhances the system from memory-less transitions to general ones, to keep track of the elapsed days spent in each disease stage. 
We provide theoretical analyses of the generation time distributions for general infectious profiles and sojourn times. As far as we know, this analysis has not been considered in previous studies of discrete-time epidemic models. In particular, we obtain the relative contributions of the asymptomatic and symptomatic states to this distribution \citep{Fraser(2004)}.

\section{The model} \label{Sec:2}

The starting point for the enhanced discrete-time epidemic model that we are going to introduce is the unstructured model introduced and analyzed in \citep{Ripoll(2023)}. Here and in the previous work, we set the time step as one day, and thus all the introduced probabilities are referred to as probabilities per day.

 The system in \citep{Ripoll(2023)} describes the spread of an infectious disease during an epidemic outbreak with asymptomatic carriers. 
 \textcolor{blue}{The model assumes a time sequence of three pivotal events: exposure to the pathogen, the onset of transmission, and subsequently, the onset of symptoms.}
 It takes the form of a non-linear Markov chain for the fraction of Susceptible $S_t$, Exposed $E_t$ (latent who are not infectious yet), infectious Asymptomatic $A_t$ (who may eventually develop symptoms), Infectious symptomatic $I_t$, Removed $R_t$ (alive and immune) and (disease-related) Deceased $D_t$ hosts, at time $t$ in days. Like in many models, the waiting times at the infected stages $E \to A \to I$, are based\footnote{The authors in \citep{Ripoll(2023)} also introduced an enhanced model based on negative binomial distributions and with reinfections due to loss of immunity.} on \textit{geometric distributions} (discrete analog of exponential distributions in continuous time). Following the notation in \citep{Ripoll(2023)}, if $X_E, X_A$ and $X_I$ are the discrete random variables representing the duration of the latent, asymptomatic, and symptomatic stages, respectively, then 
$$
\pr(X_E>k)= (1-\alpha)^k \; , \quad \pr(X_A>k)= (1-\delta)^k \; , \quad \pr(X_I>k)= (1-\gamma)^k \; , \; k \geq 1 \; \text{ days} \; ,
$$
with probabilities per day: $0<\alpha, \delta, \gamma< 1$. Accordingly, one gets the expected waiting times as follows: $\E[X_E]=\frac{1}{\alpha}$ is the mean latent period, and $\E[X_A]=\frac{1}{\delta}$ and $\E[X_I]=\frac{1}{\gamma}$ are the mean infectious period for the asymptomatic and symptomatic hosts, respectively. For the sake of completeness, let us write the epidemic model in \citep{Ripoll(2023)} as the following recursive system with geometric waiting times and variable infectiousness only across phases:

\begin{equation}\label{recurrent}
\left\{ \begin{array}{l}
     S_{t}= {\d\prod_{k=1}^{\infty}} (1-\varepsilon_{t-k} )= \exp\left(-\d\sum_{k=1}^{\infty} \frac{\beta^A A_{t-k} +\beta^I I_{t-k}}{1-D_{t-k}} \right)  \\
     E_{t}= {\d\sum_{k=1}^{\infty}} (1-\alpha)^{k-1}\varepsilon_{t-k} S_{t-k} \\
     A_{t}= \alpha {\displaystyle\sum_{k=1}^{\infty}} (1-\delta)^{k-1}E_{t-k} \\
     I_{t}= \sigma \,\delta {\displaystyle\sum_{k=1}^{\infty}} (1-\gamma)^{k-1}A_{t-k} \\
\end{array}
\right. \quad t\geq 1 \; \text{ days}
\end{equation}
where the force of infection $\varepsilon_t= 1 - e^{-(\beta^A A_t +\beta^I I_t)/(1-D_t)}$ with transmission rates $\beta^A, \beta^I >0$, depends on the infectious individuals over alive population. The fraction of Removed and Deceased hosts are
$R_t=1-(S_t+E_t+A_t+I_t+D_t)$ and $D_t= q\gamma {\d\sum_{j=1}^{\infty}} I_{t-j}$, respectively,
and $0<\sigma, q<1$ are the probability of developing symptoms after the asymptomatic phase, and the proportion of symptomatic cases that result in death (case fatality ratio), respectively. For a set of suitable discrete initial histories to system (\ref{recurrent}) and more modeling details, see sections 2 and 3 in  \citep{Ripoll(2023)}. 

These geometric/exponential waiting times, however, are not always the most suitable ones, according to reported data. In fact, the length of infectious periods has been observed to follow an empirical distribution which is often approximated by log-normal and gamma (Smallpox), fixed-length (measles), or Weibull (Ebola) distributions \citep{Chowell(2014), Ferretti(2020), Kiss(2015), Lloyd(2001)}. So, beyond memory-less waiting times, we can increase modeling flexibility by assuming general discrete probability distributions for waiting times at the infected stages: 
$$
\pr(X_E>k)= p_k^E \; , \quad \pr(X_A>k)= p_k^A \; , \quad \pr(X_I>k)= p_k^I \; , \; k \geq 1 \; \text{ days}\; .
$$
Accordingly, for each infected stage, the expected period is computed by the usual formulas for positive discrete random variables: $\E[X_i]= \displaystyle\sum_{k=1}^{\infty} k (p_{k-1}^i-p_k^i)= \sum_{k=1}^{\infty} p_{k-1}^i \geq 1$ with $p_0^i:=1$, $i=E,A,I$. See Table \ref{tab:WTprobabilities} for a summary.

\begin{table}[t]
    \centering
    \begin{tabular}{|c|p{10cm}|} \hline
         $\pr(X_E>k)=p_k^E$& probability that a host remains {latent} (\textbf{Exposed}) more than $k$ days after exposure. \smallskip\\ \hline 
         $\pr(X_A>k)=p_k^A$& probability that a host remains \textbf{infectious Asymptomatic} (mild or no symptoms) more than $k$ days after transmission onset.\smallskip\\ \hline
         $\pr(X_I>k)=p_k^I$& probability that a host remains \textbf{Infectious symptomatic} more than $k$ days after symptom onset.\smallskip \\ \hline
    \end{tabular}
    \caption{Discrete random waiting times $X_E, X_A, X_I$ for the latent (inactive), asymptomatic and symptomatic stages, respectively, characterized by specific probabilities $0\leq p_k^E, p_k^A, p_k^I\leq 1$, where index $k\geq 1$ is the number of elapsed days. $p_0^E=p_0^A= p_0^I:=1$. An asymptomatic host refers to either a pre-symptomatic individual or a life-long asymptomatic carrier. Related periods: $X_E+X_A$ is the incubation period, $X_A+X_I$ is the period of communicability, and $X_E+X_A+X_I$ is the total duration of the infection.}
    \label{tab:WTprobabilities}
\end{table}

\textcolor{blue}{
Note that, based on the previous probability distributions, the time spent in one compartment is assumed to be independent of the other sojourn times. For instance, the time an individual spends in the asymptomatic compartment is independent of the time already spent in the latent compartment: $\pr(X_A=k \, | \, X_E=l)=\pr(X_A=k)$. This lack of statistical dependence between sojourn times is an assumption present in most epidemic models, whether deterministic or based on branching processes.  
}

Next, we introduce new state variables $E_{t,j}$, $A_{t,j}$ and $I_{t,j}$, as described in Table \ref{tab:NewVariables}. 
\begin{table}[b]
    \centering
    \begin{tabular}{|c|l|} \hline 
         $E_{t,j}$& fraction of hosts at time $t$, exposed for $j$ days.\\ \hline 
         $A_{t,j}$& fraction of hosts at time $t$, infectious and asymptomatic for $j$ days.\\ \hline 
         $I_{t,j}$& fraction of hosts at time $t$, infectious and symptomatic for $j$ days.\\ \hline
    \end{tabular}
    \caption{Elapsed-time structured variables. Fraction of the host population at each infected stage $E \to A \to I$, structured by the elapsed time since the event, $j\geq 1$, where \textit{event} refers to exposure, onset of transmission, or onset of symptoms, respectively.}
    \label{tab:NewVariables}
\end{table}
Accordingly, the fraction of hosts, at any \textit{age} (i.e. the elapsed time since a specific \textit{event}) that are Exposed, infectious Asymptomatic and Infectious symptomatic are given by
$$
E_t= \sum_{j=1}^{\infty} E_{t,j} \, , \quad
A_t= \sum_{j=1}^{\infty} A_{t,j} \, , \quad
I_t= \sum_{j=1}^{\infty} I_{t,j} \, ,
$$
respectively. 
Finally, the enhanced discrete-time epidemic model with variable infectiousness and random waiting times, reads as the following non-Markovian recursive system
\begin{equation}\label{general_model}
\left\{ \begin{array}{l}
     S_{t}= {\d\prod_{k=1}^{\infty}} (1-\varepsilon_{t-k})= \exp\left(-\d\sum_{k,j=1}^{\infty} \frac{\beta_j^{A} A_{t-k,j} +\beta_j^{I} I_{t-k,j}}{1-D_{t-k}} \right)  \medskip \\
     E_{t,k}= p_{k-1}^E \cdot \varepsilon_{t-k} S_{t-k} \quad \medskip\\
     A_{t,k}= p_{k-1}^A \, \d\sum_{j=1}^{\infty} \left(1-\frac{p_{j}^E}{p_{j-1}^E} \right) E_{t-k,j}\quad   \medskip\\
     I_{t,k}=  p_{k-1}^I \, \d\sum_{j=1}^{\infty} \left(1-\frac{p_{j}^A}{p_{j-1}^A} \right) \sigma_j \cdot A_{t-k,j} \quad 
\end{array}
\right.   \quad t,k \geq 1 \; \text{ days}
\end{equation}
where $\beta_j^{A}, \beta_j^{I}\geq 0$ are transmission rates after $j$ days of transmission onset and symptom onset, respectively, and $0\leq \sigma_j \leq 1$ is the probability of developing symptoms after $j$ days of transmission onset. 
The fraction of removed and deceased hosts are $R_t= 1-(S_t+E_t+A_t+I_t+D_t)$ and 
\begin{equation} \label{Deceased} \textstyle
D_t=  {\d\sum_{k,j=1}^{\infty}}  \left(1-\frac{p_{j}^I}{p_{j-1}^I} \right) q_j \cdot I_{t-k,j}\; ,
\end{equation}
respectively, where $0\leq q_j \leq 1$ is the case fatality ratio after $j$ days of symptom onset. Here, the force of infection at time $t$ is given by:
\begin{equation}\label{force_infection}
\varepsilon_{t}=1 - \exp\left(-\d\sum_{j=1}^{\infty} \frac{\beta_j^{A} A_{t,j} +\beta_j^{I} I_{t,j}}{1-D_t} \right) 
\end{equation}
and is based on a Poisson distribution for the number of contacts per day between pathogens and susceptible hosts.
We have assumed known discrete-time histories in $(-\infty,0]$ for system (\ref{general_model}), i.e. the state variables in the past, such that
$\lim_{k \to \infty} S_{-k}=1$, and $\lim_{k \to \infty} E_{-k}= \lim_{k \to \infty} A_{-k}= \lim_{k \to \infty} I_{-k}= \lim_{k \to \infty} R_{-k}= \lim_{k \to \infty} D_{-k}= 0$.

\begin{figure}[t] 
\centering\includegraphics[width=16cm]{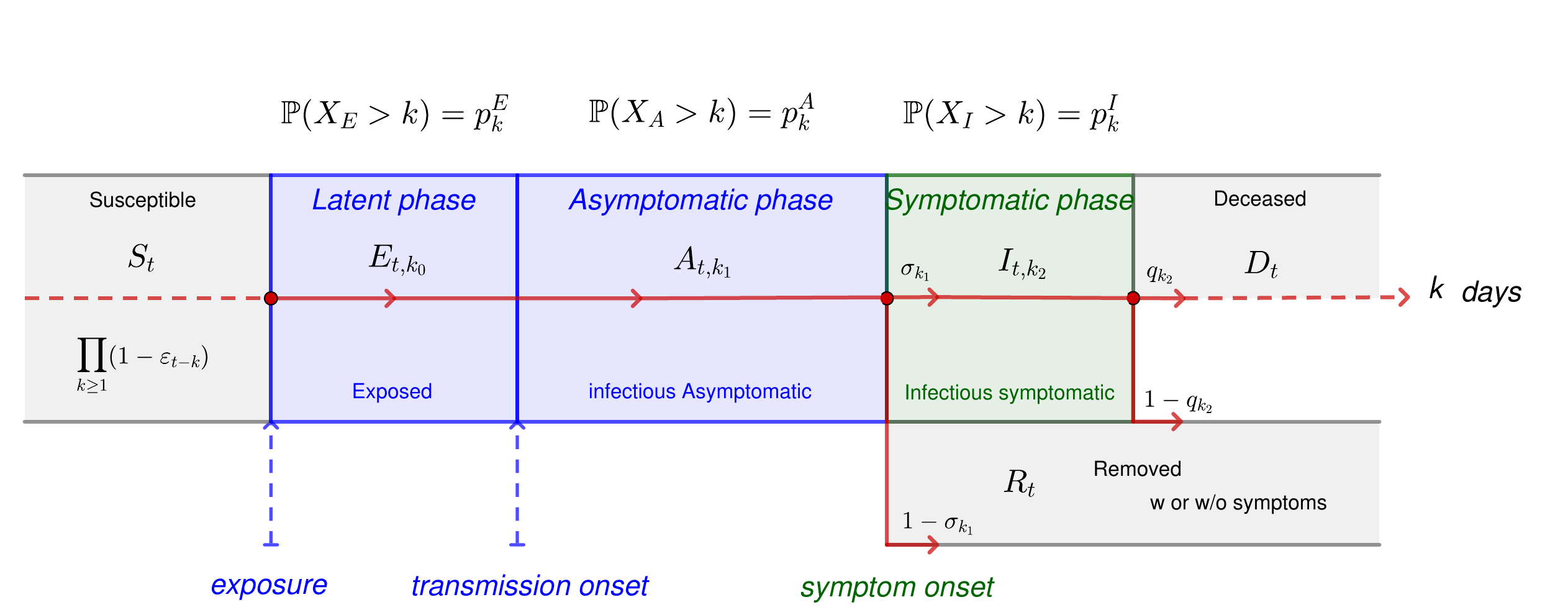}	
\caption{Flow diagram of system (\ref{general_model}), the \textit{SEA}-\textit{RID} non-linear discrete-time epidemic model with asymptomatic carriers. \textcolor{blue}{Hosts progress through the exposure to the pathogen, the onset of transmission, and subsequently, the onset of symptoms.} The discrete random variables $X_E, X_A, X_I\geq 1$ represent the waiting times at the infected stages. State variables are expressed as a fraction of the host population: Susceptible, Exposed $k_0\geq 1$ days ago, infectious Asymptomatic $k_1\geq 1$ days ago, Infectious symptomatic $k_2\geq 1$ days ago, Removed (alive and immune) and disease-related Deceased. The probability of remaining susceptible until time $t$ is $\prod_{k\geq 1} (1-\varepsilon_{t-k})$. The probabilities of remaining at stages $E$, $A$ or $I$ for more than $k$ days are $p_k^E, p_k^A, p_k^I$, respectively. Additionally, $\sigma_{k_1}$ denotes the probability of developing symptoms $k_1$ days after the onset of infectiousness, and $q_{k_2}$ represents the case fatality ratio $k_2$ days after the onset of symptoms.
See Tables \ref{tab:WTprobabilities} and \ref{tab:NewVariables}.\label{Fig:flow}}
\end{figure}


See Figure \ref{Fig:flow} for the disease progression of the epidemic model (\ref{general_model}). 

Model equations (\ref{general_model}) can be explained in words as follows. The first equation says that the fraction of susceptible hosts at time $t\geq 1$, $S_t$, is given by the probability of not being infected at any time in the past, i.e. $t-k$, $k\geq 1$. The second equation says that the fraction of exposed hosts since $k$ days at time $t$, $E_{t,k}$, is computed as the incidence (fraction of new cases per day) at time $t-k$, $\varepsilon_{t-k} S_{t-k}$, times the probability of remaining latent $k$ days or more after exposure, $p_{k-1}^E= \pr(X_E > k-1)= \pr(X_E \geq k)$. Lastly, the third and fourth equations have both a similar explanation. Namely, the fraction of infectious asymptomatic hosts since $k$ days at time $t$, $A_{t,k}$, is computed by adding over $j\geq 1$, the fraction of exposed hosts since $j$ days at time $t-k$, $E_{t-k,j}$, times the probability of becoming infectious after $j$ days from the exposure, 
$1-\frac{p_{j}^E}{p_{j-1}^E}= \frac{\pr(X_E = j)}{\pr(X_E>j-1)}$; and times the probability of remaining infectious Asymptomatic $k$ days or more after transmission onset $p_{k-1}^A$. Analogously, the last equation in (\ref{general_model}) says that the fraction of Infectious symptomatic hosts since $k$ days at time $t$, $I_{t,k}$, is computed by adding over $j\geq 1$, the fraction of infectious Asymptomatic hosts since $j$ days at time $t-k$, $A_{t-k,j}$, times the probability of becoming symptomatic after $j$ days from the transmission onset, provided that the host develops symptoms $\sigma_j \neq 0$,
$(1-\frac{p_{j}^A}{p_{j-1}^A}) \,\sigma_j= \frac{\pr(X_A = j)}{\pr(X_A>j-1)} \,\sigma_j$; and times the probability of remaining symptomatic $k$ days or more after symptom onset $p_{k-1}^I$.  
Finally, notice that the number of deceased hosts $D_t$ is simply computed in (\ref{Deceased}) by adding the number of symptomatic cases in the past taking into account the case fatality ratio, $q_j$, $j\geq 1$.

In summary, the unstructured Markovian recursive system (\ref{recurrent}) is enhanced to the infection-age structured non-Markovian recursive system (\ref{general_model}) featuring arbitrarily distributed waiting times and variable infectiousness along elapsed times and across phases. Notice that (\ref{general_model}) is non-Markovian unless the case of geometric waiting times and time-independent parameters $\beta^A, \beta^I$, $\sigma$, and $q$.

For discrete-time models in population biology, we refer to the books \citep{Otto(2007),Seno(2022),Elaydi(2025),Thieme(2024)}. See also \citep{Diekmann(2021),Brauer(2010),Hernandez(2013)}.

\section{Renewal equation and reproduction number} \label{Sec:3}

To get to the discrete renewal equation for the fraction of asymptomatic hosts we proceed as follows.
 
From the second equation in (\ref{general_model}), we have that $E_{t,k}= p_{k-1}^E \varepsilon_{t-k}\prod_{n=1}^{\infty} (1-\varepsilon_{t-k-n})$ and the third equation for $k\geq 1$ becomes
$$ 
    A_{t,k}= \d\sum_{i=1}^{\infty} p_{k-1}^A \left(1-\frac{p_{i}^E}{p_{i-1}^E} \right) E_{t-k,i} = \d\sum_{i=1}^{\infty} (p_{i-1}^E-p_{i}^E) p_{k-1}^A  \varepsilon_{t-k-i} \d{\prod_{n=1}^{\infty}} (1-\varepsilon_{t-k-i-n}) \, .
$$ 
Linearizing this equation around the disease-free steady state, $S^*=1, E^*=A^*=I^*=0$, and using that the force of infection, given by (\ref{force_infection}), is approximated by $\varepsilon_t \simeq \d\sum_{j=1}^{\infty} {\beta_j^{A} A_{t,j} +\beta_j^{I} I_{t,j}}$ around this equilibrium,
we have that
$$
    A_{t,k}= \d\sum_{i,j=1}^{\infty} (p_{i-1}^E-p_{i}^E) p_{k-1}^A  \left({\beta_j^{A} A_{t-k-i,j} +\beta_j^{I} I_{t-k-i,j}} \right) \; , \, k\geq 1 \; ,
$$
and using the last equation in (\ref{general_model}) for the symptomatic hosts $ I_{t-k-i,j}$, it becomes
\begin{equation} \label{renewal}
    A_{t,k}= \d\sum_{i,j=1}^{\infty} (p_{i-1}^E-p_{i}^E) p_{k-1}^A  \left({\beta_j^{A} A_{t-k-i,j} +\beta_j^{I}  p_{j-1}^I \, \sum_{n=1}^{\infty} \left(1-\frac{p_{n}^A}{p_{n-1}^A} \right) \sigma_n \cdot A_{t-k-i-j,n} } \right) \;, \, k\geq 1 \,  .
\end{equation}
which is the linear renewal equation for $(A_{t,k})_{k\geq 1}$, the sequence with respect to the number of days since infectiousness onset, of the fraction of asymptomatic hosts.

Either from the modeling ingredients of the present model, see Fig. \ref{Fig:flow}, or from the above discrete renewal equation (with $A_{t,k}= p_{k-1}^A$), we get to the basic reproduction number for system (\ref{general_model}), that is, the transmission potential of the infectious disease as:
\begin{equation} \label{R00}
\mathcal{R}_0= \d\sum_{i,j=1}^{\infty} (p_{i-1}^E-p_{i}^E)  \left({\beta_j^{A} p_{j-1}^A +\beta_j^{I}  p_{j-1}^I \sum_{n=1}^{\infty} \left({p_{n-1}^A}-{p_{n}^A} \right) \sigma_n} \right)
\end{equation}
which simplifies to
\begin{equation} \label{R0}
    \mathcal{R}_0= \d\sum_{j=1}^{\infty} {\beta_j^{A} p_{j-1}^A +\bar{\sigma} \cdot \beta_j^{I} p_{j-1}^I} 
\end{equation}
with the average $\bar{\sigma}:= \sum_{n= 1}^\infty \sigma_n \cdot \pr(X_A=n)>0$ being the expected probability of developing symptoms after the asymptomatic phase. Notice that index $j$ has a different meaning in $\beta_j^{A} p_{j-1}^A$ and $\bar{\sigma} \cdot \beta_j^{I} p_{j-1}^I$, which is the number of days since transmission onset and the number of days since symptom onset, respectively.

The infection event here is meant as the exposure to the pathogen of a susceptible host becoming an asymptomatic individual, so $\mathcal{R}_0$ in (\ref{R0}) is interpreted as the expected number, at the beginning of an epidemic outbreak, of secondary asymptomatic cases produced by an asymptomatic primary case who may eventually develop symptoms. See \citep{Barril(2021b)} for an analogous computation in a continuous-time model.

Moreover, the expression in (\ref{R0}) is made up by two terms $\mathcal{R}_0= \mathcal{R}_0^A +\mathcal{R}_0^I$ corresponding to the contribution of each phase to the transmission of the disease:
\begin{equation} \label{contribution}
\begin{array}{ll}
  \mathcal{R}_0^A= \sum_{j\geq 1} \beta_j^{A} p_{j-1}^A = \langle \beta^A \rangle \cdot \E[X_A] & \textbf{asymptomatic contribution} \medskip\\
  \mathcal{R}_0^I= \bar{\sigma} \sum_{j\geq 1}  \beta_j^{I} p_{j-1}^I = \bar{\sigma}\, \langle \beta^I \rangle \cdot \E[X_I]
& \textbf{symptomatic contribution} \\
\end{array}
\end{equation}
where the brackets stand for the mean transmission rate across each infectious phase: 
$$\textstyle \langle \beta^i \rangle := \big(\sum_{k\geq 1}\beta_k^{i} p_{k-1}^i\big) / \big(\sum_{k\geq 1} p_{k-1}^i \big) \; , \quad i=A,I \; ,$$ 
that is, the transmission rate taking into account the probability that a host remains infectious.
In this way, we can interpret the basic reproduction number 
\begin{equation} \label{R0Compact}
\mathcal{R}_0= \langle \beta^A \rangle \cdot \E[X_A] + \bar{\sigma} \, \langle \beta^I \rangle \cdot \E[X_I]
\end{equation}
as, on average, the number of infections during the asymptomatic phase (first term: \textit{mean transmission rate $\times$ the expected duration of the asymptomatic phase}) plus, provided that the host develops symptoms, the number of infections during the symptomatic phase (second term: \textit{mean transmission rate $\times$ the expected duration of the symptomatic phase}). 

Finally, let us remark that $\mathcal{R}_0$ in (\ref{R0}) is formally computed from the renewal equation (\ref{renewal}) on the space of sequences by the standard approach of next-generation matrix/operator (the sequence $\big(p_{k-1}^A\big)_{k\geq 1}$ of the probabilities that a host remains infectious Asymptomatic is the non-negative eigenvector corresponding to the basic reproduction number), see \citep{Barril(2017),Barril(2021a),Barril(2021b),Breda(2020),Breda(2021)} and \citep{Diekmann(1990)}.

Beyond the computation above, we can compute a time-dependent analog of the basic reproduction number (\ref{R0}) which is an indicator of the transmission potential of the infectious disease at time $t$:
$$
 \mathcal{R}_t= S_t \d\sum_{j=1}^{\infty} \frac{1-e^{-\beta_j^{A} A_{t,j}}}{A_{t,j}} p_{j-1}^A  + \bar{\sigma} \cdot  \frac{1-e^{-\beta_j^{I} I_{t,j}}}{I_{t,j}} p_{j-1}^I  \; .
$$
This expression is called \textit{effective reproduction number} and it is interpreted as the expected number of new cases produced by an asymptomatic case who may eventually develop symptoms, provided that the situation at the $t$-th day remains unchanged, \textcolor{blue}{see chapter 9 in \citep{Seno(2022)} and \citep{Ripoll(2023)}.
Actually, it is based on a Poisson distribution for the number of contacts per day, and the terms $1-e^{-\beta_j^{A} A_{t,j}}$ and $1-e^{-\beta_j^{I} I_{t,j}}$ correspond to the force of infection at the $t$-th day due to asymptomatic and symptomatic hosts at the $j$-th elapsed day, respectively, and neglecting the deceased hosts ($D_t\simeq 0$).} 

\section{Generation time}
In this section we focus on the timing of infection events.
Generation time in epidemic models relates times between \textit{infector} and \textit{infectee}, specifically, it is the elapsed time between new cases in a chain of infection transmission. Individuals who infected more than one individual generate several generation times. Of course, generation-time instances have not all the same duration, so we can consider the generation time as a positive random variable $T$ with a given probability distribution. Let us remark that generation time $T$ may differ from the serial interval $T_S$, see below, which is a related observable quantity defined as the elapsed time between the symptom onset of an \textit{infector} and the symptom onset in its \textit{infectees}, see \citep{Svensson(2007)}.

In the discrete-time setting, the generation time $T>0$ is described by its discrete probability distribution (PMF), 
$$\mathbb{P}(T= s)= \omega_s \; , \quad s \geq 1 \; \text{ days}$$ with $\sum_{s\geq 1} \omega_s=1$, and we are interested in the moments of the generation time
$\mathbb{E}[T^n]= \sum_{s\geq 1} s^n \, \omega_s$, $n\geq 1$.
Next, we are going to compute the discrete set of probabilities $\omega_s$, $s\geq 1$, i.e. the probability that the generation time is equal to $s$ days, from the model ingredients. In general, the connection is possible if we focus on the \textit{early phase of the infection} and we rewrite the expression of the basic reproduction number $\mathcal{R}_0$ given by (\ref{R00}) from terms
referred to the waiting times at the infected stages to terms referred to the timing $s\geq 1$ of the infection events.
Indeed, rearranging the terms of $\mathcal{R}_0$ in (\ref{R00}), we get
\begin{equation} \label{R0s}
    \mathcal{R}_0= \d\sum_{s=1}^{\infty} \sum_{k=1}^{s-1} (p_{k-1}^E-p_{k}^E)  \left(\beta_{s-k}^{A} \, p_{s-k-1}^A  + \d\sum_{m=1}^{s-k-1} \left(p_{m-1}^A - p_{m}^A \right) \sigma_m \, \beta_{s-k-m}^{I} \, p_{s-k-m-1}^I \right)
\end{equation}
with $\sum_{k=1}^{0}= \sum_{m=1}^{0}=\sum_{m=1}^{-1}:= 0$. Therefore, by the definition of $\omega_s$ as the probability of the infection-events timing, we get to the following explicit expression:
\begin{equation} \label{omega}
\omega_s= \frac{1}{\mathcal{R}_0} \sum_{k=1}^{s-1} (p_{k-1}^E-p_{k}^E)  \left(\beta_{s-k}^{A} \, p_{s-k-1}^A  + \d\sum_{m=1}^{s-k-1} \left(p_{m-1}^A - p_{m}^A \right) \sigma_m \, \beta_{s-k-m}^{I} \, p_{s-k-m-1}^I \right)
\end{equation} 
where $\mathcal{R}_0$ is computed from either (\ref{R0}) or (\ref{R0s}). We recall that $p_0^E= p_0^A= p_0^I:=1$ and notice that $\omega_1=0$ because latent period is at least one day, $\omega_2= \frac{1}{\mathcal{R}_0} (1-p_1^E) \beta_1^A$, etc. Therefore, the generation time is at least two days.
The expression of the basic reproduction number in (\ref{R0s}) can also be derived from the modeling ingredients of the present model, see Fig. \ref{Fig:flow}, taking waiting times $E \to A$: $k$ and $(s-k)$ days, respectively, and waiting times $E \to A \to I$: $k$, $m$ and $(s-k-m)$ days, respectively.

\begin{figure}[t] 
\centering\includegraphics[height=8cm]{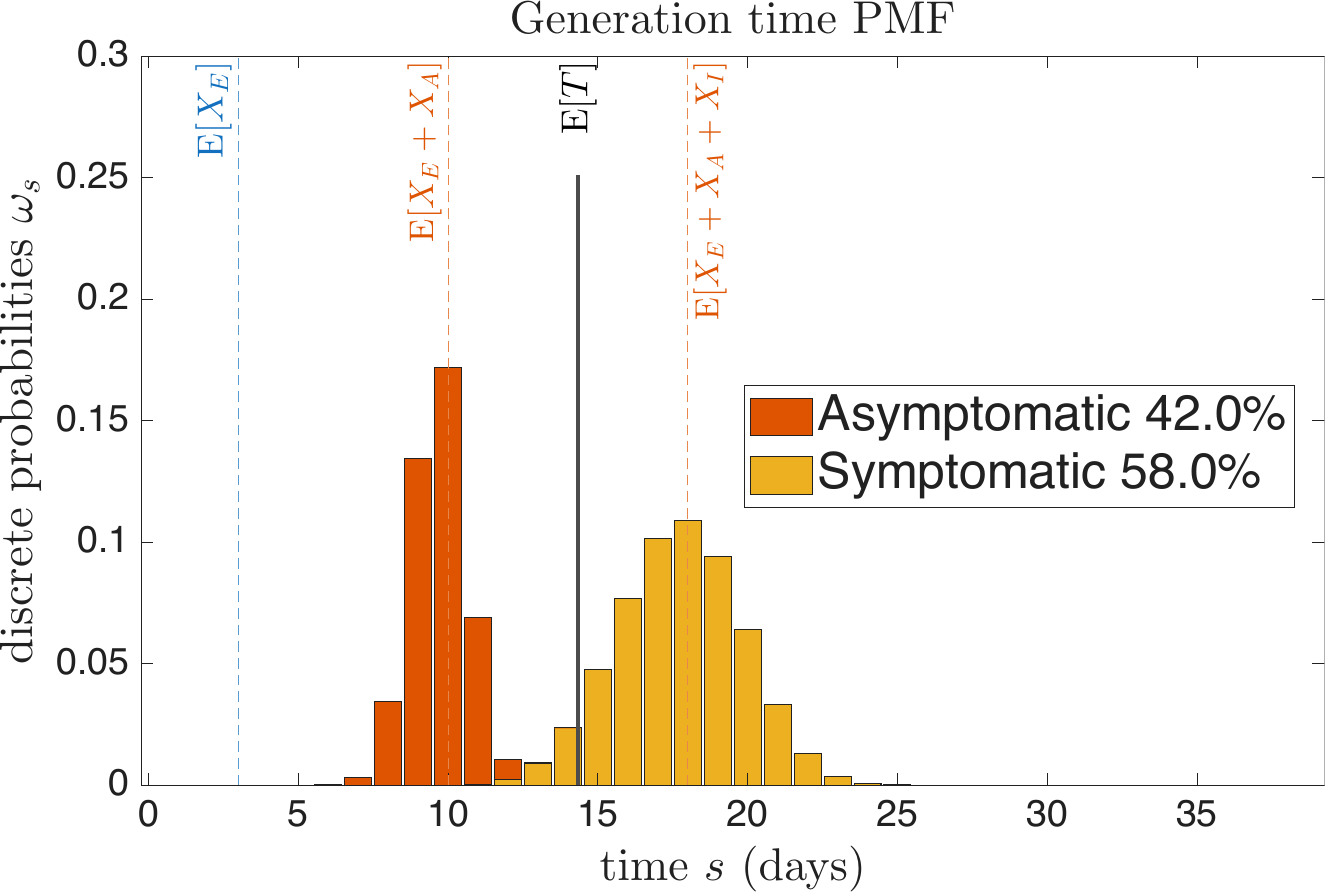}
\caption{Illustration of the generation time distribution when transmission rates $\beta^A_j$ and $\beta^I_j$ depend on the elapsed times, $j\geq 1$, and are concentrated around $\E[X_A]$ and $\E[X_I]$, respectively. This represents a generic case for a disease with two separate periods of high transmission. Each stacked bar represents $\pr(T=s)$, the probability (\ref{omega_symbols}) of the time between successive cases in a transmission chain, displaying a bimodal distribution. The latent period is $\E[X_E]=3$ days; asymptomatic infections are concentrated around $\E[X_E+X_A]=10$ days and symptomatic ones around $\E[X_E+X_A+X_I]=18$ days since exposure to the pathogen. The resulting generation time $T$ is $14.3 \pm 4.3$ days (\texttt{mean$\,\pm\,$std}). The red and yellow areas account for the respective contributions of each phase. Random waiting times $X_E, X_A, X_I$, see Table \ref{tab:WTprobabilities}, follow discrete Weibull distributions (\ref{Weibull}) with shape parameters $\alpha_E = 60$, $\alpha_A=\alpha_I=4$, and scale parameters $\theta_i =(\E[X_i]-0.5)/\Gamma(1+1/\alpha_i)$, $i=E,A,I$. Probability of developing symptoms is set to $\sigma=0.7$ . Vertical dashed lines indicate the expected lengths of one, two, or three combined stages.\label{Fig:Bimodal}}
\end{figure}

In summary, for the enhanced epidemic model (\ref{general_model}), the discrete probability distribution of the generation time $T$, in terms of the probabilities of the waiting times at infected stages, $X_E,X_A,X_I$,  
 is expressed in symbols as:
\begin{equation} \label{omega_symbols}
\mathbb{P}(T= s)= \frac{1}{\mathcal{R}_0}\sum_{k=1}^{s-1} \pr({\scriptstyle X_E=k}) \left(\beta_{s-k}^{A} \, \pr({\scriptstyle X_A \geq s-k}) + \d\sum_{m=1}^{s-k-1} \sigma_m \, \beta_{s-k-m}^{I} \, \pr({\scriptstyle X_A=m}) \pr({\scriptstyle X_I \geq s-k-m}) \right) \; , \quad s \geq 2 \; \text{ days}
\end{equation} 
where $\beta_{s-k}^{A}$ and $\beta_{s-k-m}^{I}$ are the transmission rates after infectiousness onset and symptom onset, respectively, $\sigma_m$ is the probability of developing symptoms and $\mathcal{R}_0$ is the basic reproduction number, computed either from (\ref{R0s}) or (\ref{R0Compact}).
Additionally, we can readily compute the moments of the generation time distribution as $\mathbb{E}[T^n]= \sum_{s\geq 1} s^n \, \mathbb{P}(T= s)$, $n\geq 1$. 

For simplicity in the sequel, we will assume that the conditional probability of developing symptoms is constant along the elapsed time \textcolor{blue}{in the asymptomatic compartment}, i.e. $\sigma_m= \sigma >0$, $m\geq 1$. 
\textcolor{blue}{
In other words, spending more time asymptomatic does not change the probability of developing symptoms once the asymptomatic period is over. This assumption is consistent with the idea of statistical independence between compartments assumed in the derivation of the model. It ensures that the probability $\sigma$ of transitioning from $A$ to $I$ (and consequently from $A$ to $R$) is always the same, regardless of the number of days spent as an asymptomatic individual.
}

From (\ref{omega}), the $n$-th moment is given by 
$$ 
{\E[T^n]=} \frac{1}{\mathcal{R}_0} {\d\sum_{s\geq 2}\sum_{k=1}^{s-1}} s^n \,\pr(X_E=k) \,\beta_{s-k}^{A} \, p_{s-k-1}^A  + \frac{1}{\mathcal{R}_0} {\d\sum_{s\geq 3}\sum_{k=1}^{s-1}} {\d\sum_{m=1}^{s-k-1}} s^n \,\pr(X_E=k) \pr(X_A=m)\, \sigma \,\beta_{s-k-m}^{I} \, p_{s-k-m-1}^I 
$$ 
and, changing the order of summation in the series, 
we get to
$$ 
\E[T^n]= \frac{1}{\mathcal{R}_0} {\d\sum_{j,k\geq 1}} (j+k)^n \pr(X_E=k) \, \beta_{j}^{A} \, p_{j-1}^A  +  \frac{1}{\mathcal{R}_0} {\d\sum_{j,k,m\geq 1}} (j+k+m)^n \pr(X_E=k) \pr(X_A=m) \, \sigma \, \beta_{j}^{I} \, p_{j-1}^I,
$$ 
which can be expressed as
$$ 
\E[T^n]= \frac{1}{\mathcal{R}_0} \Big({\d\sum_{j\geq 1}} \, \E[(j+X_E)^n] \, \beta_{j}^{A} \, p_{j-1}^A  + {\d\sum_{j\geq 1}} \,\E[(j+X_E+X_A)^n] \,\sigma \,\beta_{j}^{I} \, p_{j-1}^I \Big) \; .
$$ 
This expression is a convex combination of the $n$-th moment of the generation time before and after symptom onset, and can be written as follows:
\begin{equation} \label{momentsT}
\E[T^n]= \frac{\mathcal{R}_0^{A}}{\mathcal{R}_0} \, \frac{\sum_{j\geq 1}\E[(j+X_E)^n] \,\beta_j^{A} p_{j-1}^A }{\sum_{j\geq 1}\beta_j^{A} p_{j-1}^A} + \frac{\mathcal{R}_0^{I}}{\mathcal{R}_0} \frac{\sum_{j\geq 1} \E[(j+X_E+X_A)^n] \,\beta_j^{I} p_{j-1}^I }{\sum_{j\geq 1}\beta_j^{I} p_{j-1}^I},
\end{equation}
where $\mathcal{R}_0= \mathcal{R}_0^A + \mathcal{R}_0^I= \sum_{j\geq 1}\beta_j^{A} p_{j-1}^A + {\sigma} \sum_{j\geq 1}\beta_j^{I} p_{j-1}^I$. 

In words, the $n$-th moment of the generation time is related to the moments up to $n$-th order of the {weighted forward recurrence time} at each phase, i.e. $\frac{\sum_{j\geq 1} j^n \,\beta_j^{i} \,p_{j-1}^i }{\sum_{j\geq 1}\beta_j^{i} p_{j-1}^i}$, $i=A,I$, and the moments up to $n$-th order of the latent and incubation periods, $\E[X_E^n]$ and $\E[(X_E+X_A)^n]$, respectively. Notice that the weights in the forward recurrence time are the infectiousness of each phase along the elapsed times, see \ref{sec:variance}, and the convex combination takes into account the relative contribution to the transmission of each phase.

In particular, for $n=1$, the expected generation time is the convex combination of the expected generation time before and after showing symptoms:
\begin{equation} \label{E[T]simple}
\E[T]= \frac{\mathcal{R}_0^{A}}{\mathcal{R}_0} \, \frac{\sum_{j\geq 1}(j+\E[X_E]) \,\beta_j^{A} p_{j-1}^A }{\sum_{j\geq 1}\beta_j^{A} p_{j-1}^A} + \frac{\mathcal{R}_0^{I}}{\mathcal{R}_0} \frac{\sum_{j\geq 1} (j+\E[X_E+X_A]) \,\beta_j^{I} p_{j-1}^I }{\sum_{j\geq 1}\beta_j^{I} p_{j-1}^I} \; .
\end{equation}
From this expression, it is straightforward to get a lower bound as follows:
$$
\E[T]\geq  \E[X_E] + 1 + \frac{\mathcal{R}_0^{I}}{\mathcal{R}_0} \,\E[X_A] > \E[X_E] \; .
$$

Moreover, the discrete probability distribution of the generation time is a mixture distribution, see \ref{sec:variance} for the details, i.e. a convex combination of two probability distributions corresponding to the generation times before and after symptom onset. Therefore, beyond the expectation of a mixture already given in (\ref{E[T]simple}), we can readily find the variance of a mixture using the law of total variance which splits $\text{Var}(T)$ into within-phase variance and between-phase variance. See \ref{VarT} for the formula of $\operatorname{Var}(T)$ in terms of the waiting times and the weighted forward recurrence times. 

See Fig. \ref{Fig:Bimodal} for an illustration of a bimodal distribution of the generation time when infectiousness is variable along elapsed times and there are two separate periods of high transmission. 


\subsection{Particular cases}
In this section we are going to compute the expected generation time for particular cases. The aim is to illustrate the formulas of the previous sections ((\ref{R0}), (\ref{omega}) and (\ref{E[T]simple})) applied to different epidemic scenarios.

\begin{itemize}
\item 
First case, let us assume a disease such that $\langle \beta^A \rangle \to 0$ and $\bar{\sigma}>0$, so infection events take place mainly on the symptomatic phase. Here, Exposed hosts and Asymptomatic hosts are the same (i.e. indistinguishable). For simplicity, let us assume in addition that latent phase and the asymptomatic phase are of fixed length $\pr(X_E= T_E)=1 = \pr(X_A= T_A)$, $T_E,T_A \geq 1$ days. In this scenario, $\bar{\sigma}= \sigma_{T_A}$,
$\mathcal{R}_0=\mathcal{R}_0^I= \sigma_{T_A} \sum_{j\geq 1}  \beta_j^{I} p_{j-1}^I$ 
and (\ref{omega}) becomes
$$
\omega_s= \frac{ \sigma_{T_A}\beta_j^{I} p_{j-1}^I}{\mathcal{R}_0^I} \; , \quad 
s= j +T_E + T_A \; , \quad j\geq 1
$$
giving an expected generation time
$$
\E[T]= T_E + T_A + \frac{\sum_{j\geq 1}j \,\beta_j^{I} p_{j-1}^I }{\sum_{j\geq 1}\beta_j^{I} p_{j-1}^I} = \E[T_S] \qquad \text{(serial interval)}
$$
where the last term in the sum, as before in (\ref{E[T]simple}), is the average time (days) of infection events since the onset of symptoms.
 Here, the expected generation time $\E[T]$ equals to the expected serial interval $\E[T_S]$ since there is no asymptomatic transmission and the incubation period (from exposure to symptom onset) is the same for any \textit{infector} and any \textit{infectee}. 

\item Second case, let us assume a disease with transmission before and after symptoms show up and, for simplicity,
latent and asymptomatic phases are of fixed length $T_E, T_A\geq 1$ days, respectively. Therefore, the incubation period $T_E+T_A$ is the same for any \textit{infector} or \textit{infectee} too. For this pretty general case, $\bar{\sigma}= \sigma_{T_A}$, the basic reproduction number is given by
$$\mathcal{R}_0= \left(\d\sum_{j=1}^{T_A} \beta^A_{j}\right) +\sigma_{T_A} \left(\d\sum_{j= 1}^{\infty}  \beta_j^{I} p_{j-1}^I\right)$$ 
and (\ref{omega}) becomes
$$\left\{
\begin{array}{ll}
  \omega_s= 0 & s= 1, \dots , T_E\medskip \\
  \omega_s= \frac{\beta^A_{j}}{\mathcal{R}_0}   &  s=j+T_E \; , \quad j=1, \dots , T_A \medskip\\
   \omega_s= \frac{ \sigma_{T_A}\beta_j^{I} p_{j-1}^I}{\mathcal{R}_0}  &   s= j +T_E +T_A \; , \quad j \geq 1   
\end{array}\right. \;.
$$
Accordingly, we have that the expected generation time is given by
$$
\E[T]= T_E + \frac{1}{\mathcal{R}_0} \left( \d\sum_{j=1}^{T_A} j \,\beta^A_{j} + \sigma_{T_A} \sum_{j= 1}^{\infty} (j+T_A) \beta_j^{I} p_{j-1}^I \right) \; .
$$
Moreover, the variance of the generation time is computed by $\operatorname{Var}(T)=\E[T^2]-\E[T]^2$ with 
$$
\E[T^2]= \frac{1}{\mathcal{R}_0} \left( \d\sum_{j=1}^{T_A} (j+T_E)^2 \,\beta^A_{j} + \sigma_{T_A} \sum_{j= 1}^{\infty} (j+T_E+T_A)^2 \beta_j^{I} p_{j-1}^I \right) \; .
$$
 See also \ref{sec:variance} for the variance of the general case.
\end{itemize}

\subsection{Constant transmission rates} \label{sec:ct_infectivity}

In this section, we will leverage the assumption that infectiousness varies only across phases, in addition to constant probability of developing symptoms. We are going to illustrate the generation time distribution when waiting times at infected stages are given by discrete Weibull distributions (which include geometric distributions and, as a limit case, fixed-length distributions). The latter is a suitable distribution to model waiting times in many epidemic scenarios \citep{Chen(2022)}, and ecological ones, e.g. \citep{Calsina(2007)}.

\begin{figure}[h] 
\centering\includegraphics[height=8cm]{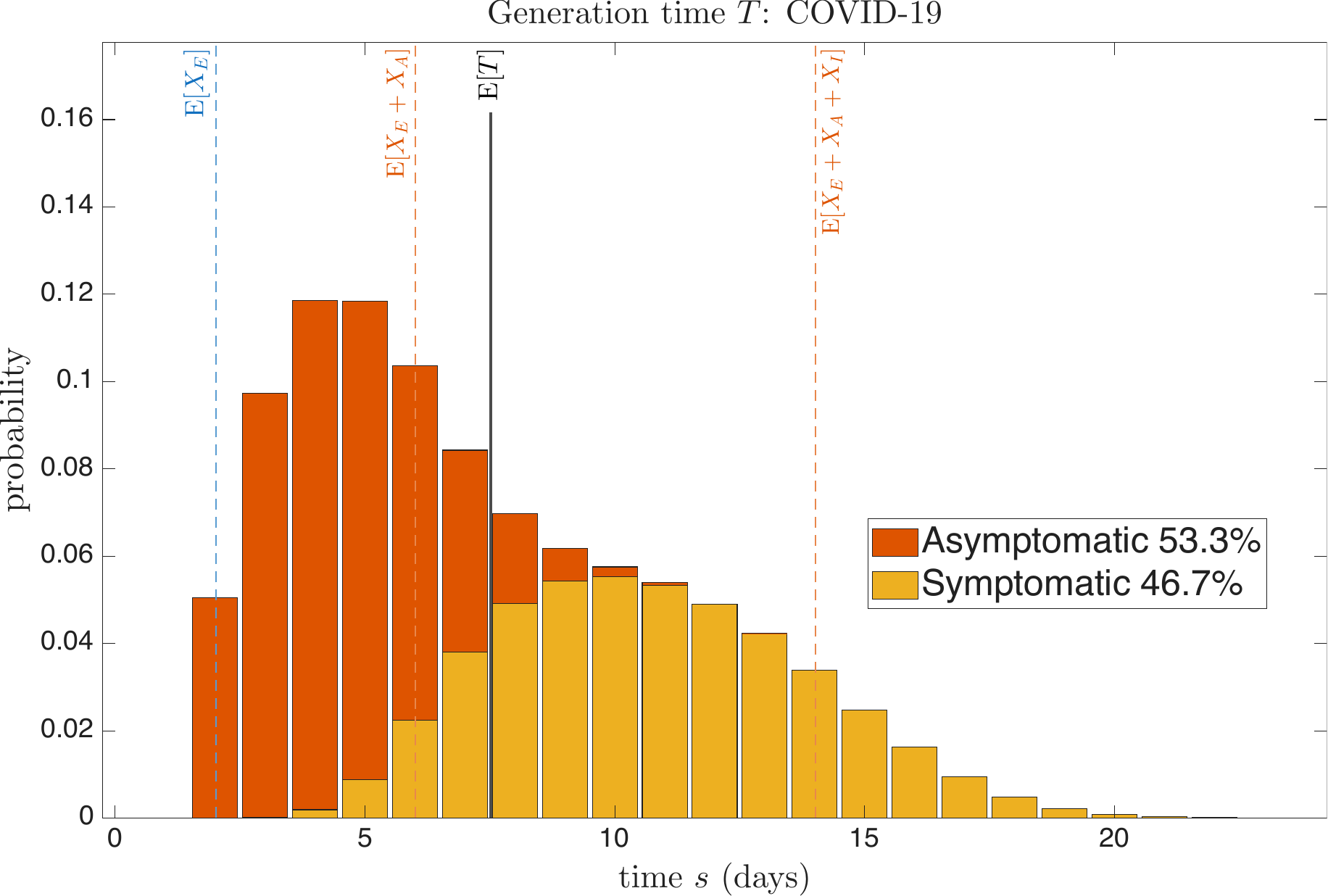}
\caption{Estimated generation time (GT) distribution for COVID-19 assuming constant transmission rates $\beta^A, \beta^I$. Each stacked bar represents the probability (\ref{omega_ct}) of the time between successive cases in a transmission chain. The contribution of asymptomatic hosts is larger than that of symptomatic ones \textcolor{blue}{resulting in a median of the GT distribution equal to the mean incubation period, $\E[X_E+X_A]$, of $6$ days. This reflects the well-known fact for COVID-19 that many transmissions occur before the incubation period is over \citep{Linton(2022)}}. Waiting times at each stage follow a Weibull distribution (see \eqref{Weibull}). Vertical dashed lines correspond, from left to right, to the expected lengths of the latent period, the incubation period, and the total duration of the infection, respectively. \textcolor{blue}{The expected generation time $\E[T]= 7.5$ days, computed from Eq.~\eqref{E[T]ctsimple}, is closer to the mean incubation period than to the mean duration of the infection, see \citep{Ferretti(2020),Chen(2022)}.} Parameter values are taken from the second row in Table \ref{tab:values}. \label{Fig:COVID-19}}
\end{figure}

If transmission rates and the probability of developing symptoms are independent of 
the elapsed times\footnote{This hypothesis corresponds to constant infectiousness along elapsed times but variable infectiousness across phases.}
, that is, 
\begin{equation} \label{constant_case}
\beta_j^{A}= \beta^{A}>0 \, , \; \beta_j^{I}=\beta^{I}>0 \; \text{ and } \; \sigma_j= \sigma >0 \, , \; j\geq 1 \, ,
\end{equation} 
then (\ref{omega}) becomes
\begin{equation} \label{omega_ct}
\omega_s= \frac{1}{\mathcal{R}_0} \sum_{k=1}^{s-1} (p_{k-1}^E-p_{k}^E)  \left(\beta^{A} \, p_{s-k-1}^A  + \sigma \, \beta^{I} \d\sum_{m=1}^{s-k-1} \left(p_{m-1}^A - p_{m}^A \right) p_{s-k-m-1}^I \right) \; , \quad s \geq 2 \; \text{ days} \; ,
\end{equation} 
and (\ref{momentsT}) becomes a simpler convex combination:
\begin{equation} \label{momentsTct}
\E[T^n]= \frac{\mathcal{R}_0^{A}}{\mathcal{R}_0} \, \frac{\sum_{j\geq 1}\E[(j+X_E)^n] \, p_{j-1}^A }{\sum_{j\geq 1} p_{j-1}^A} + \frac{\mathcal{R}_0^{I}}{\mathcal{R}_0} \frac{\sum_{j\geq 1} \E[(j+X_E+X_A)^n] \, p_{j-1}^I }{\sum_{j\geq 1} p_{j-1}^I} \; ,
\end{equation}
where $\mathcal{R}_0= \mathcal{R}_0^A + \mathcal{R}_0^I= \beta^{A}\cdot \mathbb{E}[X_A] + \sigma \,\beta^{I}  \cdot \mathbb{E}[X_I]$.
Here, the forward recurrence time at each phase is independent of transmission rates $\beta^{A},\beta^{I}$ which only appear in the relative contribution of each phase (coefficients of the convex combination). 

In particular, for $n=1$ in (\ref{momentsTct}) and (\ref{EYct}) for the expected forward recurrence time at each phase, we get to the discrete version of Svensson's formula \citep{Svensson(2007)} with two phases:
\begin{equation} \label{E[T]ctsimple}
\E[T]= \frac{\mathcal{R}_0^{A}}{\mathcal{R}_0} \left(\E[X_E] + \frac{\E[X_A^2]}{2\E[X_A]} \right)  + \frac{\mathcal{R}_0^{I}}{\mathcal{R}_0} \left(\E[X_E+ X_A] + \frac{\E[X_I^2]}{2\E[X_I]} \right)+ \frac{1}{2} \; ,
\end{equation}
and it is straightforward to get a lower bound as follows:
$$ 
\E[T]\geq  \frac{\mathcal{R}_0^{A}}{\mathcal{R}_0} \E[X_E + \frac{X_A}{2}]    + \frac{\mathcal{R}_0^{I}}{\mathcal{R}_0} \E[X_E+ X_A+\frac{X_I}{2}] + \frac{1}{2} \; .
$$ 

In summary, when infectiousness is constant along elapsed times but variable across phases, expression (\ref{E[T]ctsimple}) gives the expected generation time in terms of the expected 
duration of the infected phases $X_E, X_A, X_I$ and the variability (e.g. variance or standard deviation) of the length of the transmission stages $X_A, X_I$. The continuous counterpart of (\ref{E[T]ctsimple}) would be without the extra half day. 

In addition to the expected generation time (\ref{E[T]ctsimple}), for the specific expression of $\text{Var}(T)$ in terms of the moments up to third order of the waiting times, see formula (\ref{VarTct}) in \ref{sec:variance}.

For illustration purposes, we will assume that each waiting time is given by a discrete Weibull distribution with shape parameter $\alpha>0$ and scale parameter $\theta>0$:
\begin{equation}\label{Weibull}
\begin{array}{l}
\pr(X_i>k)= e^{-\left({k}/{\theta_i} \right)^{\alpha_i}} \, , \; k\geq 1\, ,  \medskip\\
 \E[X_i] = {\d\sum_{k\geq 0}} e^{-\left({k}/{\theta_i} \right)^{\alpha_i}}  \simeq \theta_i \cdot \Gamma(1+1/\alpha_i)+ \frac{1}{2} \geq 1\medskip\\
 \text{Var}(X_i)\simeq \theta_i^2 \cdot \Gamma(1+2/\alpha_i)- (\theta_i \cdot \Gamma(1+1/\alpha_i))^2 - \frac{1}{12}
\end{array} \qquad i=E,A, I \, .
\end{equation}
The expectation above is approximated by the expectation of the continuous Weibull distribution, that is, in general $\int_0^\infty f(x) \, dx\simeq \sum_{k \geq 0} \frac{f(k)+f(k+1)}{2}=\sum_{k \geq 0} f(k) - \frac{f(0)}{2}$, under the convergence assumption. The 1/12 in the variance above corresponds to the well-known shift between discrete and continuous variances (i.e. Sheppard's correction for grouping).
On the one hand,
for the special case of shape $\alpha_i=1$, we recover the geometric distribution with probability $p= 1-e^{-1/\theta_i}$. In this case, $\E[X_i] = \frac{1}{p} \simeq \theta_i + 1/2$ and $\text{Var}(X_i)= \frac{1-p}{p^2} \simeq \theta_i^2 -1/12$. On the other hand,
 for the limit case $\alpha_i \to \infty$, we get fixed-length distributions, i.e. $\pr(X_i=T_i) \to 1$ and $\text{Var}(X_i) \to 0$. Notice that we have high and low variance $\text{Var}(X_i)$ for shapes in $0< \alpha_i \leq 1$ and $\alpha_i>1$, respectively. Moreover, (\ref{Weibull}) defines a system to numerically compute the shape $(\alpha)$ and scale $(\theta)$ parameters of the Weibull distribution from the mean and variance $(\E[X],\text{Var}(X))$ of a given waiting time.



When $X \sim \text{Weibull}(\alpha,\theta)$, $\alpha, \theta>0$, then each expected forward recurrence time in (\ref{E[T]ctsimple}) can be approximated as follows:
$$
    \text{Average residual time} \, =
\; \;
\frac{\E[X^2]}{2\E[X]} + \frac{1}{2}
\simeq  \frac{\theta^2 \Gamma(1+2/\alpha)-1/3}{2\theta\Gamma(1+1/\alpha)+1}+1 \geq 1\; .
$$
When shape $\alpha=1$, we recover the geometric case $\frac{\E[X^2]}{2\E[X]} + \frac{1}{2}= \E[X]= (1-e^{-1/\theta})^{-1}$ (i.e. the expected duration of the phase) and when shape 
$\alpha \to \infty$, we get the fixed-length case $\frac{\E[X^2]}{2\E[X]} + \frac{1}{2}= \frac{\E[X]+1}{2}$ (i.e. half of the duration of the phase plus half day). On the other hand, when $0< \alpha < 1$ we have that $\frac{\E[X^2]}{2\E[X]} + \frac{1}{2}> \E[X]$ and so potentially exceeding the expected duration of the phase. 
\textcolor{blue}{Let us remark that we have chosen the Weibull distribution for its flexibility, but alternative distributions such as the Negative Binomial distribution, which is the discrete analog of the Gamma distribution, yield analogous distributions for the generation time, and so drawing the same qualitative conclusions.
}

To end up, let us apply the results of this section to several specific diseases. We have collected data on seven infectious diseases with asymptomatic carriers (seasonal influenza, COVID-19, poliomyelitis (polio), measles, TB in the pre-antibiotic era, Ebola\footnote{True asymptomatic and pre-symptomatic transmission is believed negligible for Ebola. For this disease, transmission begins after symptom onset, making it a unique boundary case in our set of diseases.} and rubella), extracted from reliable sources like WHO and CDC. Let us remark that we have found a great range of variability for most of the parameter values. 

In Table \ref{tab:values} we have reported data on $\mathcal{R}_0>1$, $b=\beta^A/\beta^I \leq 1$, $0<\sigma<1$, waiting times $X_E,X_A,X_I$, \texttt{mean$\pm$std}, and serial interval $T_S$. \textcolor{blue}{For rubella, we assigned a small but non-zero relative infectiousness of asymptomatic cases $b=0.1$, reflecting evidence that rubella may be transmitted by subclinical or asymptomatic individuals \citep{CDC(2024)}. For Ebola, we set $b=0$, because the available review evidence does not support asymptomatic transmission \citep{Dean(2016)}. In this formulation, the asymptomatic compartment remains epidemiologically relevant for disease progression and case ascertainment, even though it does not directly generate secondary infections. We treat the probability of developing symptoms after the asymptomatic phase, $\sigma$, as a modeling parameter since it is difficult to identify directly from the literature and is disease-specific. Where direct estimates were unavailable, we assigned plausible values consistent with the known clinical spectrum of each infection and the proportion of asymptomatic or subclinical cases. Because measles infection is classically symptomatic in almost all susceptible hosts, we set the probability of developing symptoms to $\sigma=0.9$ (conservative baseline of subclinical cases).} 

\textcolor{blue}{For seasonal influenza and COVID-19, the stage durations were chosen to match the central values reported in the cited epidemiological studies and reviews. Notably, the COVID-19 parameters are representative of the ancestral strain from the early pandemic; while later variants differ dynamically, we frame these baseline values as representative model inputs rather than unique empirical constants. For stage durations of poliomyelitis, we used a compact natural-history approximation consistent with standard descriptions of the disease: an incubation period of about 7–10 days, a minimal pre-symptomatic infectious stage of one day to satisfy the model structure, and a symptomatic stage of roughly one week, while recognizing that fecal shedding may persist beyond the symptomatic period. For measles, we interpret $X_A$ as the pre-symptomatic infectious period of about 4 days before rash onset. To remain consistent with the reported incubation period, we set $X_E$ to approximately 6 days, so that $X_E+X_A$	matches the typical 10-day time from exposure to clinical onset. For pre-antibiotic TB, we use a long-duration approximation consistent with the literature showing that active disease typically develops over months to two years. Because the literature does not support a precise decomposition into latent, presymptomatic infectious, and symptomatic infectious stages, the compartment durations are chosen as broad modeling inputs rather than direct estimates. So, the row for TB is meaningful only as an average progression model. For Ebola virus disease, we parameterized the stage durations using \citep{Velasquez(2015)} to distinguish the incubation (time from exposure to onset of any symptoms), latent (time from exposure to onset of wet symptoms), and infectious (duration of wet symptoms) periods, and used \citep{Chowell(2014)} and \citep{WHO(2025)} guidance to ensure consistency with outbreak-based timing estimates and the fact that transmission begins only after symptom onset. Rubella stage durations were selected to reflect the reported incubation period, asymptomatic fraction, and pre-/post-rash infectious period described in public-health sources.
}

Then we have computed the relative contribution of the symptomatic phase $\mathcal{R}^I_0/\mathcal{R}_0 \, 100 \%$, and
the generation time $T$, \texttt{mean$\pm$std} and \texttt{(median)}, assuming (\ref{Weibull}) and the parameter values for the seven diseases. 
In addition, we have also included the daily incidence growth rate $\lambda > 1$, which at the early phase of the infection is computed from the discrete Euler-Lotka equation $1= \mathcal{R}_0 \d\sum_{s\geq 1} \lambda^{-s} \omega_s$. This quantity measures the epidemic growth rate \citep{Ma(2020),Wallinga(2007)}, that is, how quickly the number of new cases increases each day, with measles being the fastest and TB being the slowest of the seven analyzed diseases.




\begin{landscape}
\begin{table}[h]	\textcolor{blue}{
\begin{tabular}{|l | r r r r r r | r | r | r | r|} 
\toprule
\textbf{Disease} & 
\rotatebox{90}{\shortstack{\textbf{Basic reproduction} \\ \textbf{number} $\mathcal{R}_0$}} & 
\rotatebox{90}{\shortstack{Relative infectiousness of \\ asymptomatics $b=\beta^A/\beta^I$}} & 
\rotatebox{90}{\shortstack{Probability of developing \\ symptoms $\sigma$}} & 
\rotatebox{90}{\shortstack{Non-infectious latent \\ period $X_E$ \texttt{mean$\pm$std} (days)}} & 
\rotatebox{90}{\shortstack{Infectious asymptomatic \\ period $X_A$ \texttt{mean$\pm$std} (days)}} & 
\rotatebox{90}{\shortstack{Infectious symptomatic \\ period $X_I$ mean$\pm$std (days)}} & 
\rotatebox{90}{\shortstack{Symptomatic contribution \\ to transmission $\mathcal{R}^I_0/\mathcal{R}_0$ (\%)}} & 
\rotatebox{90}{\shortstack{Serial interval $T_S$ (days)}} & 
\rotatebox{90}{\shortstack{Daily growth rate $\lambda$}} &
\rotatebox{0}{\shortstack{\textbf{Generation time} $T$\\ \texttt{mean$\pm$std (median)}\\ days}} \\
\midrule
Seasonal influenza $^a$ & 1.3 & 0.35 & 0.40 & 1.25$\pm$0.75 & 1.5$\pm$0.5 & 5$\pm$0.25 & 79.0\% & 2.2 -- 2.8 & 1.053 & 5.2 $\pm$ 2.0 \quad (5) \\
COVID-19 $^b$ & 3 & 0.57 & 0.25 & 2 $\pm$ 1 & 4 $\pm$ 1 & 8 $\pm$ 2 & 46.7\% & 3 -- 7.6 & 1.187 & 7.5 $\pm$ 3.9 \quad (6) \\
Poliomyelitis $^c$ & 1.8 & 1.00 & 0.03 & 7 $\pm$ 2 & 1 $\pm$ 0 & 6 $\pm$ 3 & 15.3\% & 7 -- 14 & 1.073 & 8.7 $\pm$ 2.8 \quad (8) \\
Measles $^d$ & 12 & 0.10 & 0.90 & 6 $\pm$ 1 & 4 $\pm$ 1 & 4 $\pm$ 1 & 90.0\% & 11.7 & 1.238 & 12.2 $\pm$ 2.3 \; (12) \\
TB pre-antibiotic $^e$ & 2.5 & 0.50 & 0.20 & 90$\pm$30 & 180$\pm$60 & 300$\pm$120 & 40.0\% & 182 -- 730 & 1.004 & 292.1$\pm$162.3 (251) \\
Ebola $^{f, \dagger}$ & 2 & 0.00 & 0.95 & 6 $\pm$ 2 & 1 $\pm$ 1 & 9 $\pm$ 4 & 100.0\% & 15.4 & 1.056 & 13.2 $\pm$ 4.5 \; (12) \\
Rubella $^g$ & 10 & 0.10 & 0.50 & 7 $\pm$ 1 & 7 $\pm$ 2 & 7 $\pm$ 4 & 83.3\% & 18.3 & 1.153 & 17.8 $\pm$ 5.0 \; (17) \\
\bottomrule
\end{tabular}
\caption{Reported data and the computed generation time $T$ for seven infectious diseases with asymptomatic carriers. The values of the parameters $\mathcal{R}_0>1$, $b=\beta^A/\beta^I$ (constant transmission rates), random waiting times at disease stages $X_E,X_A,X_I\geq 1$, \texttt{mean$\pm$std} (days), and serial interval $T_S$ (days between symptom onset in an infector and symptom onset in its infectees) belong to ranges reported in the literature. We treat $0<\sigma<1$ as a modeling parameter, because the available literature does not map consistently onto a unique disease-specific probability.
Each random waiting time $X$ is assumed to follow a discrete Weibull distribution with specific shape $\alpha>0$ and scale $\theta>0$ parameters. These are numerically computed from the estimated mean $\mathbb{E}[X]$ and variance $\text{Var}(X)$, see (\ref{Weibull}). The relative symptomatic contribution to transmission is computed by $\mathcal{R}^I_0/\mathcal{R}_0= \sigma \mathbb{E}[X_I]/(b \mathbb{E}[X_A] + \sigma \mathbb{E}[X_I])$, see (\ref{contribution}), and the generation time $T$, \texttt{mean$\,\pm\,$std (median)} days, from the discrete probabilities $\mathbb{P}(T= s)= \omega_s$ in (\ref{omega_ct}). During the early phase of the infection, the daily incidence growth rate $\lambda$ is computed from the equation $1= \mathcal{R}_0 \sum_{s\geq 1} \lambda^{-s} \omega_s$.\\
References for $\mathcal{R}_0, b$, periods $(X_E, X_A, X_I)$ and $T_S$: \\
$^a$ \citep{Biggerstaff(2014)}, \citep{Shaikh(2023)}, periods: \citep{Carrat(2008)}, serial: \citep{Vink(2014)}\\ $^b$ \citep{Alimohamadi(2020)}, \citep{McEvoy(2021)}, periods: \citep{Lauer(2020),Byrne(2020), Harrison(2024)}, serial: \citep{Alene(2021),Griffin(2020),Madewell(2023)}\\ $^c$ \citep{Yaari(2016)}, \citep{Shaikh(2023)}, periods: \citep{WHO(2025b),CDC(2026)}, estimated $T_S$\\ $^d$ \citep{Guerra(2017)}, \citep{Shaikh(2023)}, periods: \citep{CDC(2025),UKGov(2014),WHOEMRO(2026),Canada(2026)}, serial: \citep{Vink(2014)}\\ $^e$ \citep{Ma(2018)}, \citep{Shaikh(2023)}, periods: \citep{Behr(2018),Borgdorff(2011)}, serial: \citep{Ma(2020b)}\\ $^f$ \citep{Muzembo(2024)}, asymptomatic transmission is negligible $b=0$ \citep{Dean(2016)}, periods: \citep{Velasquez(2015),Chowell(2014),WHO(2025)}, serial: \citep{EbolaSR(2024)}\\ $^g$ \citep{Papadopoulos(2022)}, small asymptomatic infectiousness $b=0.1$ \citep{CDC(2024)}, periods: \citep{CDC(2025b),StatPearls(2025),CDA(2026)}, serial: \citep{Vink(2014)}\\
$^{\dagger}$ Note: Ebola virus disease represents a boundary case where transmission begins after symptom onset, yielding a 100\% symptomatic contribution to transmission. The pre-symptomatic stage $X_A$ serves purely for disease progression and structural consistency.\label{tab:values}}
}
\end{table}
\end{landscape}

\begin{figure}[h] 
\centering\includegraphics[height=9cm]{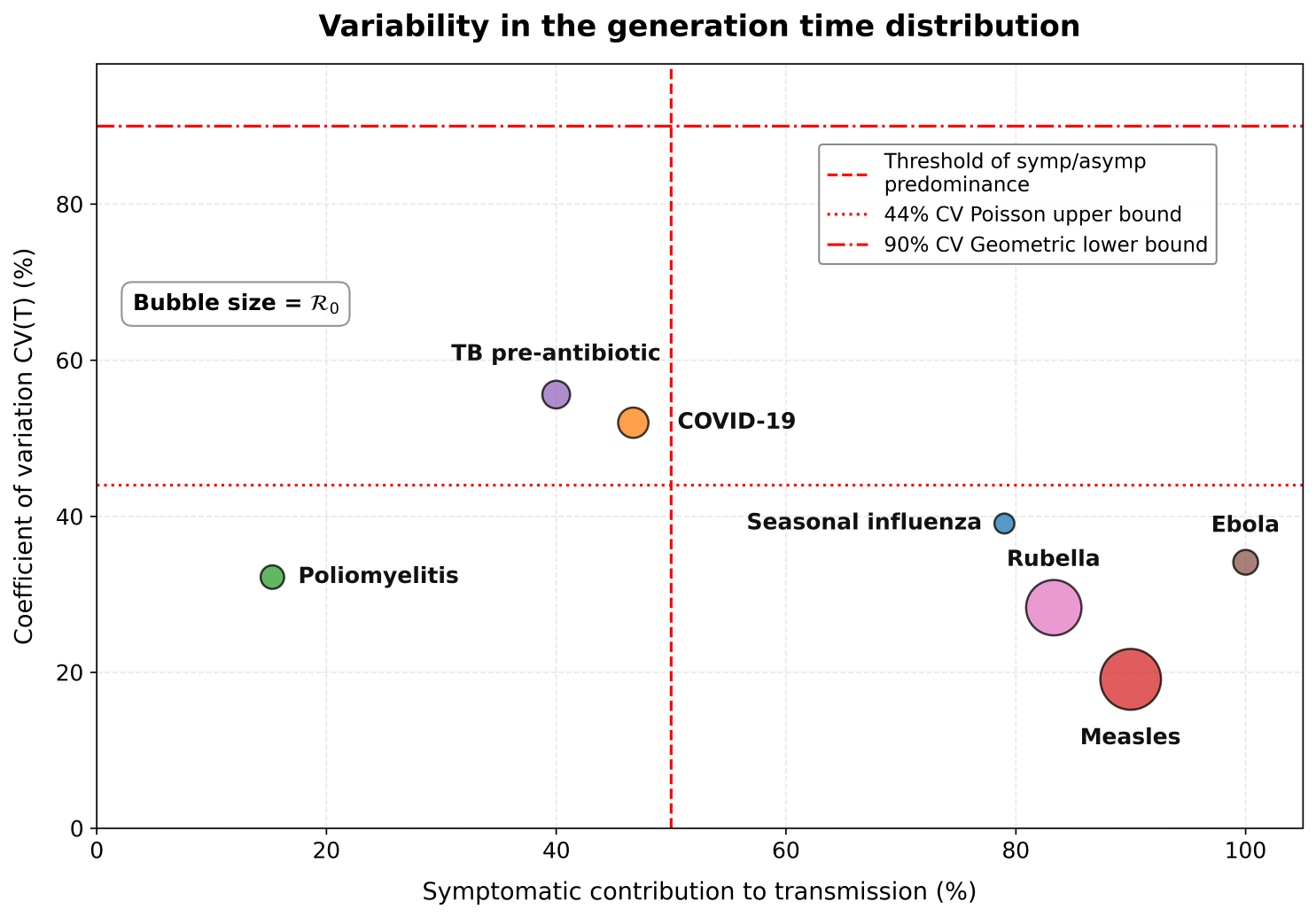}
\caption{
\textcolor{blue}{Variability in the generation time distribution across seven infectious diseases. The plot displays the coefficient of variation (CV) of the generation time as a function of the symptomatic contribution to transmission (\ref{contribution}). Bubble sizes are proportional to the basic reproduction number. Values are sourced from Table 3. Diseases situated in the left half of the plot (poliomyelitis, pre-antibiotic TB, and COVID-19) are characterized by predominant silent spread. 
The CV for all analyzed diseases falls either within a low variance range of 0\% to 44\% or within a moderate variance range of 44\% to 90\%. The 44\% threshold represents the upper bound for the CV of Poisson distributions with equivalent means ($\text{CV}=100/\!\sqrt{\text{mean}} \, \%$), whereas the 90\% threshold delineates the lower bound for geometric distributions with equivalent means ($\text{CV}=\sqrt{1-1/\text{mean}} \cdot 100\%$). \label{Fig:variability}}}
\end{figure}


According to the data displayed in Table \ref{tab:values}, infectiousness is always lower in asymptomatic hosts than in symptomatic ones, $\beta^A <\beta^I$. The probability of developing symptoms $\sigma$ is high for measles, Ebola, and COVID-19. Interestingly, the symptomatic contribution (\%) to the disease transmission is low (i.e. predominant silent spread) in poliomyelitis, pre-antibiotic TB, and COVID-19, meaning that asymptomatic infections are a major concern for those diseases.

To analyze the variability in the generation time distribution across diseases, we have summarized in Fig. \ref{Fig:variability}, the coefficient of variation $\text{CV}(T)$, the symptomatic contribution to transmission, and the basic reproduction number, for the seven analyzed diseases.

\begin{figure}[h] 
\centering\includegraphics[height=8cm]{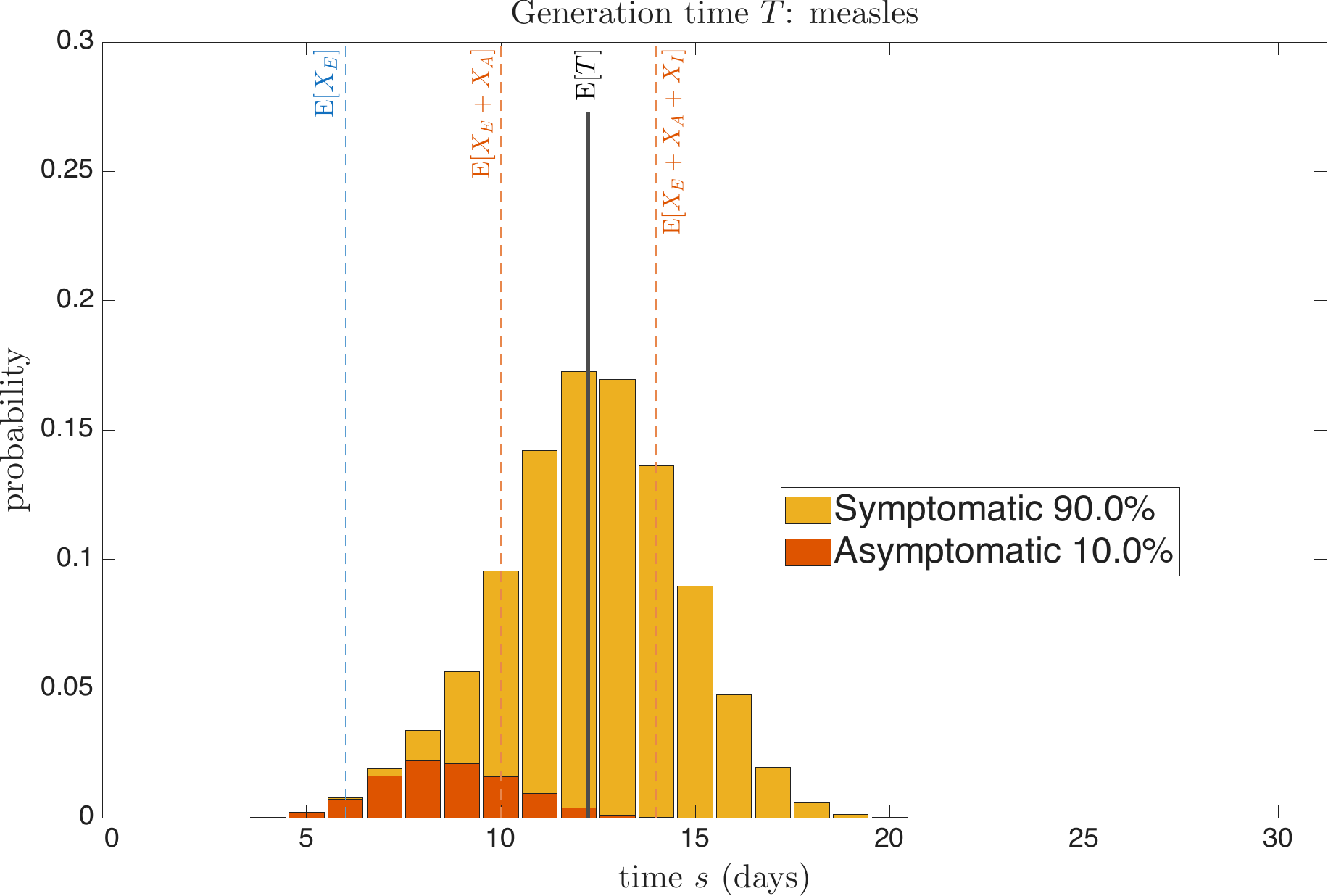 }
\caption{\textcolor{blue}{Estimated generation time (DT) distribution for measles assuming constant transmission rates $\beta^A, \beta^I$. Each stacked bar represents the probability of the timing of infection events (\ref{omega_ct}). The mean generation time, $\E[T]$, lies quite symmetrically between the expected lengths of the incubation period, $\E[X_E+X_A]$, and the total duration of the infection, $\E[X_E+X_A+X_I]$. The median ($12$ days) is almost equal to the average generation time ($12.2$ days). Waiting times follow a Weibull distribution (\ref{Weibull}). Parameter values are taken from the fourth row in Table \ref{tab:values}. \label{Fig:measles}}}
\end{figure}

\begin{figure}[h] 
\centering\includegraphics[height=8cm]{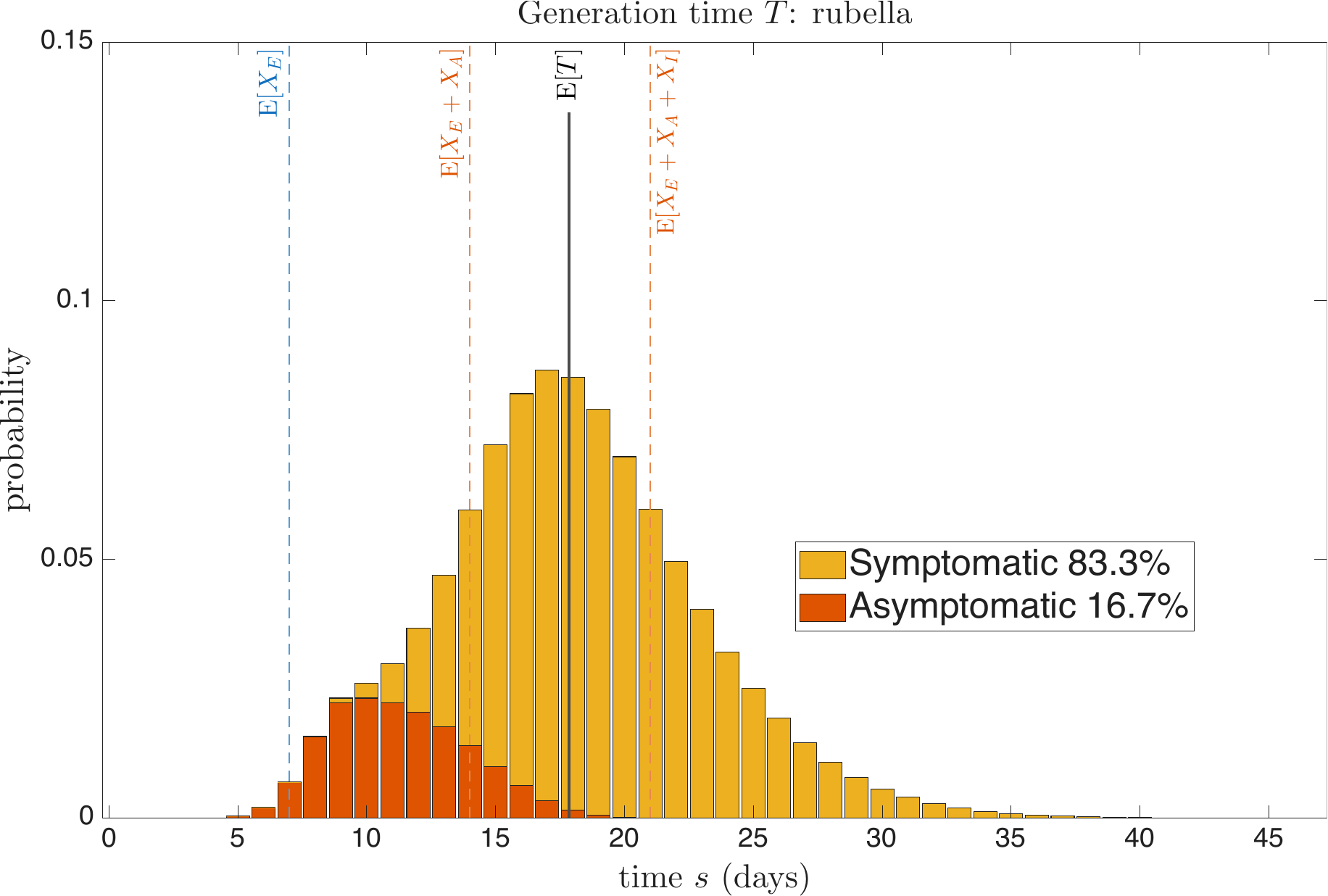}
\caption{\textcolor{blue}{Estimated generation time distribution for rubella assuming constant transmission rates $\beta^A, \beta^I$. Each stacked bar represents the probability (\ref{omega_ct}) of the time between successive cases in a chain of transmission. Despite asymptomatic individuals contributing more to transmission than in the case of measles (see
Fig.~\ref{Fig:measles}), the mean generation time, $\E[T]$, is still located quite symmetrically between the expected lengths of the incubation period, $\E[X_E+X_A]$, and the total duration of the infection, $\E[X_E+X_A+X_I]$. The median ($17$ days) is close to, but smaller than, the average generation time ($17.8$ days), reflecting a slightly right-skewed distribution. Waiting times follow Weibull distributions (\ref{Weibull}). Parameter values are taken from the last row in Table \ref{tab:values}. \label{Fig:rubella}}}
\end{figure}

Moreover, we have numerically illustrated the discrete probability distribution of the generation time for different epidemic scenarios. See Fig. \ref{Fig:COVID-19}, \ref{Fig:measles} and \ref{Fig:rubella} for the generation time of COVID-19, measles and rubella, respectively. As seen in these three plots and in Table \ref{tab:values}, the expected generation time $\E[T]$ (vertical solid black line in the plots) always lies between the expected incubation period $\E[X_E+X_A]$ and the expected total duration of the infection $\E[X_E+X_A+X_I]$ (vertical red dashed lines in the plots). \textcolor{blue}{This empirical finding does not contradict the fact that, according to formula (\ref{E[T]ctsimple}), which relates the expected times of a disease, the generation time can theoretically be shorter than the incubation period or longer than the total duration of the infection.} {In the Supplementary Material we have included the illustrations of the generation time distributions for the seven major infectious diseases considered in Table \ref{tab:values}.}



Additionally, we can compare the computed expected generation time for the seven diseases, last column in Table 
\ref{tab:values}, with the expected generation time
corresponding to the case of waiting times at transmission stages $i=A, I$, given by either geometric distributions or fixed-length distributions. From (\ref{E[T]ctsimple}), these cases are $\E[T_{\text{geom}}]= \E[X_E+X_A] + \frac{\mathcal{R}_0^{I}}{\mathcal{R}_0} \E[X_I]$ and
$\E[T_{\text{fix}}]= \E[X_E+\frac{X_A+1}{2}] + \frac{\mathcal{R}_0^{I}}{\mathcal{R}_0} \E[\frac{X_A+X_I}{2}]$, respectively. In all cases, we observe that $\E[T_{\text{geom}}]$ overestimates significantly $\E[T]$, because waiting times at disease stages are far from being memory-less, while $\E[T_{\text{fix}}]$ underestimates $\E[T]$ as expected.
Moreover, since
$$
\begin{array}{c}
\E[X_E+X_A] \leq \E[T_{\text{geom}}] \leq \E[X_E+X_A+X_I] \qquad (\text{Markovian disease stages})\medskip\\
\E[X_E+\frac{X_A+1}{2}] \leq \E[T_{\text{fix}}] \leq  \E[X_E+ X_A +\frac{X_I+1}{2}] \qquad (\text{Deterministic disease stages})
\end{array}
$$
the expected generation time for these two particular cases cannot be greater than the expected total duration of the infection.

\textcolor{blue}{Finally, to evaluate the robustness of our numerical estimates (Table \ref{tab:values} and Fig. \ref{Fig:variability}), we performed a local univariate sensitivity analysis. We perturbed the expected duration of the symptomatic infectious period, $\mathbb{E}[X_I]$, over a range of $-20\%$ to $+20\%$ in $5\%$ increments while holding its standard deviation, $\text{Std}(X_I)$, constant. For each perturbed baseline value, we numerically re-estimated the discrete Weibull shape and scale parameters and recomputed the discrete convolutions to quantify the downstream impact on the relative symptomatic contribution to transmission ($\mathcal{R}^I_0/\mathcal{R}_0$), the expected generation time $\mathbb{E}[T]$, its standard deviation $\text{Std}(T)$, and its coefficient of variation $\text{CV}(T)$.}

\textcolor{blue}{As summarized in Table \ref{tab:sensitivity}, shifting the mean symptomatic period predictably alters relative transmission weights; longer symptomatic durations proportionally increase the symptomatic contribution to overall transmission. Importantly, the core metrics governing the scale and shape of the generation time distribution, namely $\mathbb{E}[T]$, $\text{Std}(T)$, and $\text{CV}(T)$, scale smoothly and monotonically across the entire perturbation range, confirming that the overall distribution exhibits structural stability and low sensitivity to variations in stage duration.}

\begin{table}[htbp]
\centering
\textcolor{blue}{
\begin{tabular}{l c r r r r} 
\toprule
$\Delta \mathbb{E}[X_I]$  & $\mathbb{E}[X_I]$ (days) & $\Delta (\mathcal{R}^I_0/\mathcal{R}_0)$ (\%) & $\Delta \mathbb{E}[T]$ (\%) & $\Delta \text{Std}(T)$ (\%) & $\Delta \text{CV}(T)$ (\%) \\
\midrule
-20\% & 6.40 & -11.75\% & -8.51\% & -11.76\% & -3.71\% \\
-15\% & 6.80 & -8.60\% & -6.38\% & -8.95\% & -2.73\% \\
-10\% & 7.20 & -5.59\% & -4.39\% & -5.88\% & -1.79\% \\
-5\%  & 7.60 & -2.74\% & -2.13\% & -3.07\% & -0.86\% \\
\textbf{0\% (Baseline)} & \textbf{8.00} & \textbf{0.00\%} & \textbf{0.00\%} & \textbf{0.00\%} & \textbf{0.00\%} \\
+5\%  & 8.40 & +2.61\% & +2.13\% & +3.07\% & +0.85\% \\
+10\% & 8.80 & +5.09\% & +4.39\% & +6.14\% & +1.63\% \\
+15\% & 9.20 & +7.47\% & +6.65\% & +9.21\% & +2.40\% \\
+20\% & 9.60 & +9.74\% & +8.78\% & +12.28\% & +3.13\% \\
\bottomrule 
\end{tabular}
\caption{Univariate sensitivity analysis of COVID-19 generation time metrics. The table reports the impact of perturbing the mean duration of the symptomatic period, $\mathbb{E}[X_I]$, by $\pm 20\%$ in 5\% increments while holding its standard deviation $\text{Std}(X_I)$ constant. Columns 3 through 6 display the relative percentage changes relative to the unperturbed baseline (0\%).\label{tab:sensitivity}}}
\end{table}

\section{Discussion \& Conclusions} \label{Sec:8}


We have introduced a discrete-time epidemic model with variable infectiousness along elapsed times and across phases of the transmission. This model includes a latent stage, an \textit{asymptomatic but infectious stage} in which hosts show up mild or no symptoms, and a \textit{symptomatic stage} in which individuals exhibit observable clinical signs, for instance, cough, rash, or high fever. Asymptomatic cases may either develop symptoms or recover asymptomatically. The compact non-Markovian model (\ref{general_model}) is a recursive system, namely, the current state variables rely on its own past values, and features a generic random waiting time at each of the three phases: latent/exposed $E$, asymptomatic $A$ and symptomatic $I$, see Table \ref{tab:WTprobabilities} and Fig. \ref{Fig:flow}. Reported data suggest that the duration of disease stages do not follow geometric/exponential distributions in general, so we assume waiting times at stages as discrete Weibull distributions (\ref{Weibull}), gaining flexibility in the model.

\textcolor{blue}{One limitation of the model is that, although it considers general waiting time distributions in each compartment, these are assumed to be independent of each other
($\pr(X_i = a \, | \, X_j = b)=\pr(X_i=a)$ for $i \ne j$). Therefore, the length of the infectious period for an asymptomatic individual is not affected by the length of its latent period, and the same applies to the duration of the infectious period for a symptomatic individual. This \textit{independence assumption} is common to most epidemic models, whether deterministic or based on branching processes. Indeed, estimates of the SARS-CoV-2 generation time have even been derived under the assumption that the incubation period and infectiousness are independent of each other \citep{Lehtinen(2021)}.}

\textcolor{blue}{{
Other limitations of the model are: 1) the \textit{absence of age structure} (age groups) in the population, 2) the assumption of \textit{homogeneous contact patterns}, and 3) there is \textit{no waning immunity}. Finally, to simplify the expressions following Eq. \eqref{omega_symbols}, an additional hypothesis is introduced, namely, that the probability $\sigma$ of developing symptoms once the asymptomatic period is over does not change with the elapsed time in the asymptomatic compartment. This assumption aligns with the idea of \textit{statistical independence between compartments}, as it ensures that the probability of transitioning from $A$ to $I$ (and consequently from $A$ to $R$) remains constant, regardless of how many days are spent in the asymptomatic compartment.
}}

With the aim of studying the probability distribution of the generation time $T$ (the interval between being infected and infecting someone else), we firstly compute the basic reproduction number $\mathcal{R}_0$ from the asymptomatic point of view, Eq. \eqref{R0}. Specifically, the latter is
understood in the sense of the expected number of secondary cases produced by a primary case who may eventually develop symptoms. Rearranging the terms of $\mathcal{R}_0$, see Eq.\eqref{R0s}, we get to the discrete probability distribution of the generation time (\ref{omega_symbols}), depending on the model ingredients.

What we have learned is that, \textcolor{blue}{thanks to the statistical independence of the sojourn times,} the expected generation time $\E[T]$, idem for the higher-order moments $\E[T^n]$, can be computed from the generation time before symptoms and after symptoms, separately, and then combine both according to the relative contribution to transmission of each phase (convex linear combination), see Eq.~\eqref{E[T]simple} which is expressed in words as:
$$
\begin{bmatrix}
   \text{Generation}\\
    \text{time} 
\end{bmatrix} =
\Big(1-\frac{\scriptstyle\mathcal{R}_0^{I}}{\scriptstyle\mathcal{R}_0}\Big)
\begin{bmatrix}
   \text{Generation time}\\
    \text{before symptoms} 
\end{bmatrix}+
\frac{\scriptstyle\mathcal{R}_0^{I}}{\scriptstyle\mathcal{R}_0}
\begin{bmatrix}
   \text{Generation time}\\
    \text{after symptoms} 
\end{bmatrix} \; .
$$
The importance of the proportion of the transmission that occurs before the onset of symptoms or via asymptomatic transmission, has been already highlighted in \citep{Fraser(2004)}.
Equivalently, the expected generation time can be expressed in terms of expected duration of latent and asymptomatic phases, and the average time of infection events since either transmission onset or symptom onset, 
taking into account the variable infectiousness along elapsed times. Indeed, expanding formula (\ref{E[T]simple}), we can say that:
$$
\begin{bmatrix}
   \text{Generation}\\
    \text{time} 
\end{bmatrix} =
\begin{bmatrix}
   \text{Latent}\\
    \text{phase} 
\end{bmatrix} + 
\Big(1-\frac{\scriptstyle\mathcal{R}_0^{I}}{\scriptstyle\mathcal{R}_0}\Big)
\begin{bmatrix}
   \text{Timing of infections}\\
    \text{since transmission} 
    \\ \text{onset, before symptoms} 
\end{bmatrix}+
\frac{\scriptstyle\mathcal{R}_0^{I}}{\scriptstyle\mathcal{R}_0}
\,\left[\begin{matrix}
   \text{Asymptomatic}\\
    \text{phase} 
\end{matrix}+
\begin{matrix}
   \text{Timing of infections}\\
    \text{since symptom} 
    \\ \text{onset}
\end{matrix}\right] \; ,
$$
where by timing of infections we refer to a weighted forward recurrence time or residual time, see \ref{sec:variance}. So, we have computed the expected generation time in a systematic way.


Alternatively and from the modeling point of view, we could have readily reached to the expression for the expected generation time from the one of the basic reproduction number $\mathcal{R}_0$ \textcolor{blue}{if waiting times are considered mutually independent and the probability of developing symptoms $\sigma>0$ is independent of the elapsed times}. Since $\mathcal{R}_0$ is given by the sum of transmission rates times the probabilities of remaining infectious at each phase, namely,
$$
  \mathcal{R}_0= \d\sum_{j=1}^{\infty} {\beta_j^{A} p_{j-1}^A +{\sigma} \cdot \beta_j^{I} p_{j-1}^I} \; ,
$$
and index $j\geq 1$ counts the number of days since either transmission onset or symptom onset, we can heuristically derive the expected generation time as
$$
  \E[T] = \frac{1}{\mathcal{R}_0} \d\sum_{j=1}^{\infty} {\E[j+X_E] \, \beta_j^{A} p_{j-1}^A +{\sigma} \cdot \E[j+X_E+X_A] \,\beta_j^{I} p_{j-1}^I} \; ,
$$
and the $n$-th moment of the generation time distribution as
$$
  \E[T^n] = \frac{1}{\mathcal{R}_0} \d\sum_{j=1}^{\infty} {\E[(j+X_E)^n] \, \beta_j^{A} p_{j-1}^A +{\sigma} \cdot \E[(j+X_E+X_A)^n] \,\beta_j^{I} p_{j-1}^I}  \; , \quad n\geq 1 \; .
$$
Here, $X_E$ is the random number of days from exposure to transmission onset ($\E[X_E]$ is the expected latent period), and $X_E+X_A$ is the random number of days from exposure to symptom onset ($\E[X_E+X_A]$ is the expected incubation period).

Let us highlight that these \textit{direct} formulas, based on the latent or incubation period shift, are possible due to the assumption of mutually independent disease stages, and effortlessly, we would get the same results in continuous time, just changing sums $\sum_{j=1}^{\infty}$ by integrals $\int_0^\infty \, d\tau$.



\textcolor{blue}{Knowledge of the higher-order moments of the generation time distribution is useful for estimating certain epidemiological quantities when the timing of transmission is highly uncertain. For instance, they can be used to estimate the ideal length of the reporting interval for an epidemic when the variance-to-mean ratio of the generation time distribution is large \citep{Nishiura(2010b)}. Recently, upper and lower bounds for $\mathcal{R}_0$ were derived by \cite{Cochran(2024)} in terms of the first three moments of the generation time distribution.}   

If we restrict to the case of variable infectiousness only across phases, we get simpler formulas for the generation time, see Section \ref{sec:ct_infectivity}. In particular, formula (\ref{E[T]ctsimple}) is the two-phases discrete version of the Svensson's formula for the expected generation time. Here, the average time of infection events since the onset of transmission/symptoms is related to the second moment (i.e. variability) of the random waiting time at the corresponding transmission stage, see (\ref{E[T]ctsimple}).

In a similar way, if we assume that the situation at the $t$-th day remains unchanged, then from the {effective reproduction number} 
$\mathcal{R}_t= S_t  {\d\sum_{j=1}^{\infty}} \frac{1-e^{-\beta_j^{A} A_{t,j}}}{A_{t,j}} \, p_{j-1}^A + {\sigma} \cdot  \frac{1-e^{-\beta_j^{I} I_{t,j}}}{I_{t,j}} \, p_{j-1}^I$, see the end of Section \ref{Sec:3}, which is interpreted as 
the expected number of new cases produced by a case who may eventually develop symptoms, we can heuristically derive the expected {realized generation time} as 
$$\textstyle
 \quad \E[T_t]= \frac{S_t}{\mathcal{R}_t}  {\d\sum_{j=1}^{\infty}} (j+\E[X_E]) \frac{1-e^{-\beta_j^{A} A_{t,j}}}{A_{t,j}} p_{j-1}^A + {\sigma} \cdot (j+\E[X_E+X_A])\frac{1-e^{-\beta_j^{I} I_{t,j}}}{I_{t,j}} p_{j-1}^I \; .
$$
Here, $T_t$ is the time-varying random variable of the realized generation time, i.e. the elapsed time between \textit{infector} and \textit{infectee} at the $t$-th day of the ongoing outbreak, and $\E[T_t]$ depends on the state variables $S_t, A_{t,j}$ and $I_{t,j}$ \citep{Nishiura(2010)}. \textcolor{blue}{Let us note that the expression for $\mathcal{R}_t$ presented above can be formally derived in a manner similar to that outlined in chapter 9 of \citep{Seno(2022)}, and that the force of infection, which represents the daily probability of a host becoming infected, is based on a Poisson probability distribution for the daily number of contacts. Furthermore, the useful extension of the generation time, at the beginning of an outbreak, to the generation time during the outbreak, follows the same extension of the basic reproduction number to the realized reproduction number, under the assumption of independent disease stages and elapsed times $j\geq 1$ shifted by either latent or incubation period.}
{As highlighted in \citep{Hart(2022)}, empirical household contact-tracing studies show that generation times are shortened by both local susceptible depletion and non-pharmaceutical interventions, such as rapid testing, case isolation, and contact tracing.}





The generation time distributions obtained from Weibull distributions with their parameters estimated using the means and standard deviations displayed in Table~\ref{tab:values}, show a range of variability from low to moderate (see Fig.~\ref{Fig:variability}). For the seven diseases, the resulting CV is clearly smaller than that of a geometric distribution. In particular, \textcolor{blue}{the CV for measles ($18.9\%$) is the smallest one, clearly smaller than that of a Poisson distribution with the same mean ($28.6\%$)}, which is consistent with the low variability of the sojourn times for this infectious disease. Interestingly, the resulting generation time distribution for measles aligns quite well with the low-amplitude epidemic outbreaks with periods of 2 or 3 weeks reported in \citep{Keeling(1997)}. These minor measles outbreaks occur because transmission events are concentrated around the expected generation time. In our scenario, this expected value is 12.2 days, with transmission being negligible before 5 days and after 20 days (see Fig~\ref{Fig:measles}).  Generation time distributions are also related to stochastic extinctions of an infectious disease. The latter typically occur in small populations after a major epidemic peak, when there has been a significant reduction in both the number of cases and the number of susceptible individuals (epidemic fadeout). In \citep{Yang(2023)}, the dependence of epidemic fadeouts on population size was reported for a model with waning immunity, with different mean generation times and the same value of $\mathcal{R}_0$. Estimations of the intrinsic and realized generation time distributions from SARS-CoV-2 contact tracing data for the Alpha and Delta variants can be found in \citep{Manica(2022)}.  

\textcolor{blue}{If the incubation period depends on processes that affect infectiousness, then the hypothesis of statistical independence of the sojourn times will not be met, and the predicted generation time distribution will change. This would happen if both symptom onset and infectiousness depend on viral load. In this scenario, diseases that cause a faster increase in viral load are expected to have a shorter incubation period and an earlier transmission of the virus after infection \citep{Lehtinen(2021)}. This results in generation times that are smaller than those expected under the assumption that the infectious period is independent of the incubation period, that is, if $\pr(X_I=k \, | \, X_E+X_A=l)=\pr(X_I=k)$}.

Our results on the variance of the generation time distribution could help to analyze its impact on the epidemic dynamics. 
\textcolor{blue}{
For instance, low levels of variability in generation times are important for disease persistence, as epidemic fadeouts are less likely when fewer people delay transmission \citep{Keeling(1997),Lloyd(2005)}. On the other hand, variability in generation times increases the variance of the distribution of $\mathcal{R}_0$, thereby increasing the uncertainty associated with its estimate.}

\textcolor{blue}{{
A widely used approach for estimating $\mathcal{R}_0$ is the one based on the relationship between the initial exponential growth rate $r$ ($\lambda=e^r$) of the number of new cases and the generation time distribution through the Euler–Lotka equation \citep{Wallinga(2007)}. Under this approach, the uncertainty of the estimate of $\mathcal{R}_0$ arises from two sources: the estimation of growth rate $r$ and the estimation of the generation time distribution. To account for uncertainty in the generation time distribution, it is straightforward to derive a relationship between $\ln \mathcal{R}_0$ and the moments of the generation time distribution from the Taylor expansion about $z=0$ of the cumulant generating function $K(z)$ \citep{Nishiura(2010b)}. The latter is defined as the natural logarithm of the moment generating function, $K(z) = \ln \E[e^{zT}]$. Precisely, since $1/\mathcal{R}_0 = \E[e^{-rT}]$ and the $n$-th cumulant is $\kappa_n= \frac{d^n K(z)}{d z^n}\big|_{z=0}$, it follows that 
$$
\ln \mathcal{R}_0 = -K(-r) = \kappa_1 r - \frac{1}{2} \kappa_2 r^2 + \frac{1}{6} \kappa_3 r^3 - \dots
$$
with $\kappa_1=\E[T]$, $\kappa_2=\text{Var}(T)$, and the third cumulant $\kappa_3 = \E[(T-\E[T])^3]$. 
If the variability of the generation time distribution is low, the mean generation time, $\E[T]$, provides a good summary of the distribution. Neglecting second- and higher-order terms in the previous expansion then yields the well-known approximation $\mathcal{R}_0 \simeq e^{r \E[T]}$ \citep{Roberts(2007)}. 
When the generation time distribution exhibits greater variability, as is the case for some diseases considered in this paper, additional terms in the above expansion must be retained to obtain a more accurate estimate of $\mathcal{R}_0$, making bounds for this estimate increasingly important \citep{Cochran(2024)}.}}

\textcolor{blue}{ 
Greater variability in the generation time distribution also implies increased uncertainty in the timing of transmission. Consequently, public health interventions that depend on transmission timing, such as the design of contact tracing windows, should be based on a high percentile of the distribution rather than on its mean to avoid missing numerous late transmission events that occur when generation time distributions are heterogeneous.
}

\textcolor{blue}{ 
In addition to increasing the uncertainty in the estimation of $\mathcal{R}_0$, the substantial contribution of asymptomatic and pre-symptomatic individuals to the transmission of certain diseases poses additional challenges for public health interventions. This phenomenon, known as \textit{silent spread}, shifts intervention priorities from identifying symptomatic cases to interrupting transmission regardless of symptom status \citep{Weitz(2024)}. For such diseases, isolating symptomatic cases alone is insufficient to prevent many secondary infections and, hence, the epidemic spread \citep{Fraser(2004)}. Consequently, testing and contact tracing strategies must rely on proactive screening rather than symptom-based case detection, as exposed contacts need to be identified and quarantined before they develop symptoms. Testing becomes not merely diagnostic but preventive.} 

\section*{CRediT authorship contribution statement}

\textbf{Jordi Ripoll:} Writing – original draft, Writing – review \& editing, Visualization, Methodology, Investigation, Formal analysis, Data curation, Conceptualization. \textbf{Joan Saldaña:} Conceptualization, Writing - original draft, Writing – review \& editing, Funding acquisition, Methodology.

\section*{Declaration of competing interest}
The authors declare that they have no known competing financial interests or personal relationships that could have appeared to influence the work reported in this paper.

\section*{Acknowledgments}

JR and JS are members of the Catalan research group 2021 SGR 00113. This research was funded by Ministerio de Ciencia e Innovación grant numbers PID2024-157720NB-I00, PID2025-173091NB-I00, and RED2022-134784-T.
The authors would like to thank two anonymous reviewers for their valuable comments and suggestions, which greatly helped to improve the quality of this manuscript.

\appendix

\section{Reduction to a single renewal equation}

 In this appendix we are going to show the reduction of the general model (\ref{general_model}) to a single renewal equation, just adding an extra simplifying assumption on the transmission rate $\beta^A$ for the asymptomatic hosts, while keeping variable infectiousness $\beta_j^{I}$ for the symptomatic ones.
 
From both facts that $A_{t,k+j}= \beta_j^A A_{t-k,j} \frac{\pr(X_A \geq k+j)}{\pr(X_A \geq j)}$ and $A_{t,k+j}= \beta_k^A A_{t-j,k}  \frac{\pr(X_A \geq k+j)}{\pr(X_A \geq k)}$, and under the assumption that $\beta_j^A= \beta^A = \beta_k^A$, $j,k\geq 1$, we derive the useful relationship:
$$
A_{t-k,j} \cdot p_{k-1}^A = A_{t-j,k}\cdot p_{j-1}^A 
$$
to simplify the renewal equation (\ref{renewal}) to
$$ 
    A_{t,k}= \d\sum_{i,j=1}^{\infty} (p_{i-1}^E-p_{i}^E)  \left({\beta^{A} A_{t-i-j,k}\cdot p_{j-1}^A +\beta_j^{I}  p_{j-1}^I \sum_{n=1}^{\infty} \left({p_{n-1}^A}-{p_{n}^A} \right) \sigma_n \cdot A_{t-i-j-n,k} } \right) \; .
$$ 
Finally, summing with respect to $k\geq 1$ we end up with
$$ 
       A_t= \d\sum_{i,j=1}^{\infty} (p_{i-1}^E-p_{i}^E)  \left({\beta^{A} p_{j-1}^A  A_{t-i-j}+\beta_j^{I}  p_{j-1}^I \sum_{n=1}^{\infty} \left({p_{n-1}^A}-{p_{n}^A} \right) \sigma_n  A_{t-i-j-n} } \right) \; ,
$$ 
which is the (scalar) linear renewal equation for $A_t$, the fraction of infectious asymptomatic hosts. This single equation summarizes the whole system (\ref{general_model}) when linearized around the disease-free equilibrium.

\section{Back to geometric waiting times}

For the sake of completeness, let us write down the probabilities of the generation time for system (\ref{recurrent}) with memory-less waiting times, see also \citep{Ripoll(2023)}. Indeed, using the probabilities $p_k^E=(1-\alpha)^k$, $p_k^A=(1-\delta)^k$, $p_k^I= (1-\gamma)^k$, corresponding to the length of infected stages geometrically distributed, (\ref{omega_ct}) becomes
$$
\omega_s= \frac{1}{\mathcal{R}_0} \sum_{k=1}^{s-1} \alpha (1-\alpha)^{k-1}  \left(\beta^{A} \, (1-\delta)^{s-k-1}  + \sigma \, \beta^{I}\d\sum_{m=1}^{s-k-1} \delta (1-\delta)^{m-1}  (1-\gamma)^{s-k-m-1} \right) \; , \quad s \geq 2 \; \text{ days}
$$
or equivalently
$$
\omega_s=  \sum_{k=1}^{s-1} \alpha (1-\alpha)^{k-1}  \left(\frac{\mathcal{R}_0^A}{\mathcal{R}_0} \, \delta(1-\delta)^{s-k-1}  + \frac{\mathcal{R}_0^I}{\mathcal{R}_0} \d\sum_{m=1}^{s-k-1} \delta (1-\delta)^{m-1} \gamma(1-\gamma)^{s-k-m-1} \right) \; , \quad s \geq 2 \; \text{ days}
$$
with $\mathcal{R}_0= \mathcal{R}_0^A + \mathcal{R}_0^I= \frac{\beta^{A}}{\delta}+ \frac{\sigma \beta^{I}}{\gamma}$.
The expected generation time for model (\ref{recurrent}) is
$\E[T]= \left(\frac{1}{\alpha} + \frac{1}{\delta}\right) + \frac{\mathcal{R}_0^{I}}{\mathcal{R}_0}\frac{1}{\gamma}$,
which depends on the expected duration of infected stages (latent, asymptomatic and symptomatic) and the relative symptomatic contribution to transmission. Moreover, the variance of the generation time for model (\ref{recurrent}) is
$$\text{Var}(T)=  \left(\frac{1-\alpha}{\alpha^2} + \frac{1-\delta}{\delta^2}\right) + \frac{\mathcal{R}_0^{I}}{\mathcal{R}_0} \frac{1-\gamma}{\gamma^2}+ \left(1-\frac{\mathcal{R}_0^{I}}{\mathcal{R}_0}\right)\frac{\mathcal{R}_0^{I}}{\mathcal{R}_0} \frac{1}{\gamma^2} \; ,$$
which follows from the fact that for the geometric case, the waiting time at a stage and the forward recurrence time (residual time) have the same distribution, see also \ref{sec:variance}.


\section{Variance of the generation time} \label{sec:variance}
Let us recall the discrete-time dynamics of the incidence at the early phase of the infection. The incidence $i(t)$ at time $t$, i.e. the number of new cases per day, is determined by the discrete renewal equation
$
\d i(t)= \mathcal{R}_0 \sum_{s=1}^{\infty} i(t-s) \cdot \omega(s)
$
where $\omega(s):= \pr(T=s)$, $s\geq 1$, is the discrete probability distribution (PMF) of the generation time $T$. The latter is defined as the random time from infection to onward transmission, and $\mathcal{R}_0$ is the basic reproduction number (transmission potential of the disease).
For convenience we use functional notation instead of subscripts (e.g. $\omega(s)=\omega_s$) in this appendix.

Let us recall the definition of the discrete convolution of two functions $f(s),g(s)$, $s\geq 1$:
$$
(f*g)(s)= \sum_{k=1}^{s-1} f(k) \cdot g(s-k) \; ,
$$
and consider the probability distribution of the generation time given in (\ref{omega_symbols}) when the probability of developing symptoms is constant along elapsed times ($\sigma_m= \sigma >0$).
Next, we introduce the following four discrete probability distributions, related to $X_E, X_A, X_I$, the random waiting times at latent, asymptomatic, symptomatic phases:
$$
\begin{array}{ll}
  f_E(s)=\pr(X_E=s)   & , \quad g_A(s)= \frac{\beta^A(s)}{\mathcal{R}_0^A} \, \pr(X_A\geq s)=: \pr(Y_A = s) \medskip\\
  f_A(s)=\pr(X_A=s)   & , \quad g_I(s)= \frac{\sigma \, \beta^I(s)}{\mathcal{R}_0^I} \, \pr(X_I\geq s)=: \pr(Y_I = s)
\end{array} \; , \qquad s\geq 1 \; \text{ days}
$$
with normalizations $\mathcal{R}_0^A= \sum_{s\geq 1} \beta^A(s) \, \pr(X_A\geq s)$ and $\mathcal{R}_0^I= \sigma \sum_{s\geq 1}  \beta^I(s) \, \pr(X_I\geq s)$.
Notice that we have defined two new positive discrete random variables $Y_A, Y_I$ (weighted forward recurrence time or residual time) corresponding to the random time of infection events since the onset of transmission and the onset of symptoms, respectively.

Then, the discrete probability distribution for the generation
time (\ref{omega_symbols}) can be written as the following mixture distribution:
$$
\omega(s)= \frac{\mathcal{R}_0^A}{\mathcal{R}_0} (f_E * g_A)(s)+ \frac{\mathcal{R}_0^I}{\mathcal{R}_0} (f_E * f_A * g_I)(s)
$$
with $\mathcal{R}_0= \mathcal{R}_0^A+\mathcal{R}_0^I$ being the basic reproduction number. Therefore, as it is well-known, the expectation and variance of a mixture can be readily computed. Indeed, on the one hand, the expectation of the generation time is
$$
\E[T]= \frac{\mathcal{R}_0^A}{\mathcal{R}_0} \E[X_E +Y_A]+\frac{\mathcal{R}_0^I}{\mathcal{R}_0} \E[X_E +X_A +Y_I]
$$
which corresponds to Eq.~(\ref{E[T]simple}) for variable infectiousness, and to Eq.~(\ref{E[T]ctsimple}) for constant infectiousness along elapsed times. On the other hand, the variance of the generation time is given by the law of total variance (variance of a mixture distribution):
\begin{equation} \label{VarT}
\text{Var}(T)= \underbrace{\frac{\mathcal{R}_0^A}{\mathcal{R}_0} \text{Var}(X_E +Y_A)+\frac{\mathcal{R}_0^I}{\mathcal{R}_0} \text{Var}(X_E +X_A +Y_I)}_{\text{Within-phase variance}}+ \underbrace{\frac{\mathcal{R}_0^{A}\, \mathcal{R}_0^{I}}{\mathcal{R}_0^2}\mathbb{E}[Y_A-(X_A+Y_I)]^2}_{\text{Between-phase variance}} \; ,
\end{equation}
namely, the variance of each infectious phase (asymptomatic and symptomatic) plus the variance between phases. Using the independence of the random variables involved, we have that
$\text{Var}(X_E +Y_A)= \text{Var}(X_E) +\text{Var}(Y_A)$, $\text{Var}(X_E +X_A +Y_I)= \text{Var}(X_E) + \text{Var}(X_A) + \text{Var}(Y_I)$, and according to the notation in Table \ref{tab:WTprobabilities}
\begin{equation} \label{VarY}
\text{Var}(Y_i)= \frac{\sum_{j\geq 1}j^2 \,\beta_j^{i} p_{j-1}^i }{\sum_{j\geq 1}\beta_j^{i} p_{j-1}^i} - \left( \frac{\sum_{j\geq 1}j \,\beta_j^{i} p_{j-1}^i }{\sum_{j\geq 1}\beta_j^{i} p_{j-1}^i} \right)^2 \; , \quad i=A, I
\end{equation}
and 
\begin{equation} \label{EY}
\E[Y_A]= \frac{\sum_{j\geq 1}j \,\beta_j^{A} p_{j-1}^A }{\sum_{j\geq 1}\beta_j^{A} p_{j-1}^A} \quad , \qquad \E[X_A+ Y_I]= \frac{\sum_{j\geq 1} (j+\E[X_A]) \,\beta_j^{I} p_{j-1}^I }{\sum_{j\geq 1}\beta_j^{I} p_{j-1}^I} \; .
\end{equation}

Moreover, analogously to the formulas for the first moment of a positive discrete random variable, we can compute the second moment as 
$\E[X^2]= \d\sum_{j=1}^{\infty} j^2 (p_{j-1}-p_j)= \sum_{j=1}^{\infty} (2j-1)p_{j-1} \geq 1$, and the third moment  as 
$\E[X^3]= \d\sum_{j=1}^{\infty} j^3 (p_{j-1}-p_j)= \sum_{j=1}^{\infty} (3j^2-3j+1)p_{j-1} \geq 1$.
Thus, when infectiousness is constant along elapsed times $\beta^{i}_j= \beta^{i}>0$, the variance (\ref{VarY}) becomes
\begin{equation}\label{VarYct}
\text{Var}(Y_i)= \frac{\E[X_i^3]}{3\E[X_i]} - \left(\frac{\E[X_i^2]}{2\E[X_i]}\right)^2 - \frac{1}{12} \; , \quad i=A, I \; ,
\end{equation}
and the expectation (\ref{EY}) becomes 
\begin{equation}\label{EYct}
\E[Y_i]= \frac{\E[X_i^2]}{2\E[X_i]} + \frac{1}{2} \; , \quad i=A, I \; .
\end{equation}
As it is well-known, if the waiting time $X_i$, $i=A, I$, is geometrically distributed then the forward recurrence time $Y_i$ has the same distribution, and so   
$\text{Var}(Y_i)= \text{Var}(X_i)$ and $E[Y_i]= \E[X_i]$. Instead, if $X_i$, $i=A, I$, is of the fixed length $T_i$, then $Y_i$ is the discrete uniform distribution in the time interval $[1,T_i]$, and so
$\text{Var}(Y_i)= (\E[X_i]^2-1)/12$ and $E[Y_i]= (\E[X_i]+1)/2$. The continuous counterpart of (\ref{VarYct}) and (\ref{EYct}) would be without the additional one twelfth day and one half day, respectively.

To end up this appendix, let us write down the formula for the variance of the generation time (\ref{VarT}) when infectiousness is independent of elapsed times:
\begin{equation}\label{VarTct}
\begin{array}{l}
    \operatorname{Var}(T) = \frac{\mathcal{R}_0^{A}}{\mathcal{R}_0} \left(\operatorname{Var}(X_E)+ {\textstyle\frac{\E[X_A^3]}{3\E[X_A]} - \left(\frac{\E[X_A^2]}{2\E[X_A]}\right)^2}\right)+ \frac{\mathcal{R}_0^{I}}{\mathcal{R}_0} \left(\operatorname{Var}(X_E)+ \operatorname{Var}(X_A)+{\textstyle\frac{\E[X_I^3]}{3\E[X_I]} - \left(\frac{\E[X_I^2]}{2\E[X_I]}\right)^2}\right)+ \\ \qquad \qquad \; +\frac{\mathcal{R}_0^{A}\, \mathcal{R}_0^{I}}{\mathcal{R}_0^2} \left( \frac{\E[X_A^2]}{2\E[X_A]}-\E[X_A]-\frac{\E[X_I^2]}{2\E[X_I]} \right)^2-\frac{1}{12}
    \geq \operatorname{Var}(X_E) + \frac{\mathcal{R}_0^{I}}{\mathcal{R}_0} \operatorname{Var}(X_A)
    \end{array}
\end{equation}
Again, the continuous counterpart of (\ref{VarTct}) would be without the extra one twelfth day. 





\end{document}